\documentclass{aa}
\usepackage[varg]{txfonts}
\usepackage{multirow}
\usepackage{natbib,twoopt}
\usepackage{amsmath}
\usepackage{mathtools}
\usepackage{xr}
\usepackage[breaklinks=true]{hyperref} 
\bibpunct{(}{)}{;}{a}{}{,}             
\makeatletter
  \newcommandtwoopt{\citeads}[3][][]{\href{http://adsabs.harvard.edu/abs/#3}%
    {\def\hyper@linkstart##1##2{}%
     \let\hyper@linkend\@empty\citealp[#1][#2]{#3}}}
  \newcommandtwoopt{\citepads}[3][][]{\href{http://adsabs.harvard.edu/abs/#3}%
    {\def\hyper@linkstart##1##2{}%
     \let\hyper@linkend\@empty\citep[#1][#2]{#3}}}
  \newcommandtwoopt{\citetads}[3][][]{\href{http://adsabs.harvard.edu/abs/#3}%
    {\def\hyper@linkstart##1##2{}%
     \let\hyper@linkend\@empty\citet[#1][#2]{#3}}}
  \newcommandtwoopt{\citeyearads}[3][][]%
    {\href{http://adsabs.harvard.edu/abs/#3}
    {\def\hyper@linkstart##1##2{}%
     \let\hyper@linkend\@empty\citeyear[#1][#2]{#3}}}
\usepackage{etoolbox}
\makeatletter
\patchcmd\@combinedblfloats{\box\@outputbox}{\unvbox\@outputbox}{}{%
    \errmessage{\noexpand\@combinedblfloats could not be patched}%
}%
\makeatother

\begin{document}
\title{Dust opacity variations in the pre-stellar core L1544}

\author{A.~Chac\'on-Tanarro \inst{1}
\and J.~E.~Pineda \inst{1}
\and P.~Caselli \inst{1}
\and L.~Bizzocchi \inst{1}
\and R.~A.~Gutermuth \inst{2}
\and B.~S.~Mason \inst{3}
\and A.~I.~G\'omez-Ruiz \inst{4}
\and J. Harju \inst{1,5}
\and M.~Devlin \inst{6}
\and S.~R.~Dicker \inst{6}
\and T.~Mroczkowski \inst{7}
\and C.~E.~Romero \inst{6}
\and J.~Sievers \inst{8}
\and S.~Stanchfield \inst{6}
\and S.~Offner \inst{9}
\and D.~S\'anchez-Arg\"uelles \inst{10}
}
\institute{Max-Planck-Instit\"{u}t f\"{u}r extraterrestrische Physik, Giessenbachstrasse 1, 85748 Garching, Germany
\and Department of Astronomy, University of Massachusetts, Amherst, MA 01003, USA
\and National Radio Astronomy Observatory, 520 Edgemont Road, Charlottesville, VA 22903, USA
\and CONACYT-Instituto Nacional de Astrof\'isica, \'Optica y Electr\'onica, Luis E. Erro 1, 72840 Tonantzintla, Puebla, M\'exico
\and Department of Physics, P.O. Box 64, 00014 University of Helsinki, Finland
\and Department of Physics and Astronomy, University of Pennsylvania, 209 South 33rd Street, Philadelphia, PA, 19104, USA
\and ESO--European Organization for Astronomical Research in the Southern hemisphere, Karl-Schwarzschild-Str. 2, D-85748 Garching b. M\"unchen, Germany
\and School of Chemistry and Physics, University of KwaZulu-Natal, Private Bag X54001, Durban 4000, South Africa
\and Department of Astronomy, University of Texas at Austin, USA
\and Instituto Nacional de Astrof\'isica, \'Optica y Electr\'onica (INAOE), Luis Enrique Erro 1, 72840, Puebla, Mexico
}
\date{Received - / Accepted -}

\abstract{The study of dust emission at millimeter wavelengths is important to shed light on the dust properties and physical structure of pre-stellar cores, the initial conditions in the process of star and planet formation.}{Using two new continuum facilities, AzTEC at the LMT and MUSTANG-2 at the GBO, we aim to detect changes in the optical properties of dust grains as a function of radius for the well-known pre-stellar core L1544.}{We determine the emission profiles at 1.1 and 3.3 mm and examine whether they can be reproduced in terms of the current best physical models for L1544. We also make use of various tools to determine the radial distributions of the density, temperature, and the dust opacity in a self-consistent manner.}{We find that our observations cannot be reproduced without invoking opacity variations. With the new data, new temperature and density profiles, as well as opacity variations across the core, have been derived. The opacity changes are consistent with the expected variations between uncoagulated bare grains, toward the outer regions of the core, and grains with thick ice mantles, toward the core center. A simple analytical grain growth model predicts the presence of grains of $\sim$3-4 $\mu$m within the central 2\,000\,au for the new density profile.}{}


\maketitle

\section{Introduction}

Pre-stellar cores are starless dense ($n_{\mathrm{H}_2} > 10^5$ cm$^{-3}$) and cold ($T < 10 K$) self-gravitating cloud cores. They are formed from molecular clouds material, and present clear signs of contraction and chemical evolution \citepads{2005ApJ...619..379C}. Since they are considered the initial conditions of  star formation \citepads{2007ARA&A..45..339B, 2012A&ARv..20...56C}, these cores are important to understand the future evolution of protostars and protoplanetary disks \citepads[e.g. ]{2016MNRAS.460.2050Z}. 

In this work we focus on the pre-stellar core L1544, which is placed in the Taurus Molecular Cloud, 140 pc away. It is a proto-typical example of a pre-stellar core, whose physical structure has been studied in the past. First, \citetads{2002ApJ...569..815T} determined a density profile for the source using continuum emission at 1.3 mm, and a constant temperature of 8.75 K across the core. Later, \citetads{2007A&A...470..221C}, using high resolution ammonia interferometric observations, found a drop in temperature towards the center of the core. This implied that the density profile found by \citetads{2002ApJ...569..815T} needed to be modified. Finally,  \citetads{2015MNRAS.446.3731K} were able to reproduce molecular line and continuum observations of L1544 by describing it as a Bonnor-Ebert sphere \citepads{1956MNRAS.116..351B} in quasi-equilibrium contraction. However, \citetads{2015MNRAS.446.3731K} found that for reproducing the temperature drop measured by \citetads{2007A&A...470..221C}, they needed to increase the dust opacity. This could be an indication of the presence of fluffy grains, and therefore these results suggested that L1544 was the perfect target to study dust grain coagulation in pre-stellar cores. 

The opacity, $\kappa_{\nu}$, is a measurement of the dust absorption cross sections weighted by the mass of the gas and dust. It depends on the frequency, $\nu$, and physical properties of the dust grains, such as composition, mass and size: if dust grains are much smaller than the wavelength at which they are observed, $\kappa_{\nu}$ depends on the mass of the grains; if they are larger than the observing wavelength, $\kappa_{\nu}$ decreases with the grain size. However, dust grains with size similar to the wavelength become very efficient radiators, increasing the opacity up to 10 times its typical value  \citepads{1994A&A...288..929K}. Usually, matching the data with models requires an assumption regarding the opacity, and this translates into an assumption regarding the dust grain distribution and properties across the cloud. 

Usually, in pre-stellar cores the values used for the opacities are taken from grain coagulation models from \citetads{1994A&A...291..943O}. Deviations from these values, as well as their dependence on temperature, are a matter of current debate. Recent laboratory work from  \citetads{2017A&A...606A..50D} showed that for the temperature range typical for pre-stellar cores (10-30 K), the opacity does not depend on the temperature, although \citepads{1996ApJ...462.1026A} detected an inverse relation for very low temperatures (<20 K). The opacity can be approximated by a power law at millimeter wavelengths, $\kappa_{\nu} \propto \nu^{\beta}$, where $\beta$ is the spectral index of the dust. As shown by \citetads{2017A&A...606A..50D}, for example, the power law slope (or $\beta$) seems to vary depending on the frequency range. Together with the possibility of grain coagulation in dense cores, these results indicate that many factors can modify the emission observed.

Observationally, previous studies have found variations in the opacity and the spectral index from molecular clouds to clumps and cores \citepads[see e.g.,]{2016A&A...588A..30S, 2016ApJ...826...95C, 2015A&A...584A..94J,2015A&A...584A..93J, 2013MNRAS.428.1606F}. Going to smaller scales, \citetads{2015A&A...580A.114F} found opacity variations when studying the extinction map of the starless core FeSt 1-457, while \citetads{2017A&A...604A..52B} and \citetads{2017A&A...606A.142C} found no evidence of opacity and spectral index variations towards pre-stellar cores using the NIKA camera at the IRAM 30 m telescope.. The observed variations can be attributed to grain growth towards dense clumps and cores, although other factors, such as the noise of the data or temperature dependence should be taken into account \citepads{2009ApJ...696..676S, 2009ApJ...696.2234S}. 
In these dense regions, ice mantle growth and grain coagulation is expected to affect the emission of dust grains at submillimeter and milimeter wavelengths \citepads{2009A&A...502..845O,2011A&A...532A..43O}. Thus, the opacity behavior in dense, cold pre-stellar cores remains debated and highly uncertain.

Here we present sensitive continuum maps observed with two new facilities for millimeter observations:  AzTEC at the Large Millimeter Telescope Alfonso Serrano (LMT), observing at 1.1 mm, and MUSTANG-2 at the Green Bank Observatory (GBO), observing at 3.3 mm.  These maps help us to understand the physical structure of the core, including new information on the behavior of the opacity towards L1544. This work presents the first continuum map of L1544 at 3.3 mm, and it improves the work done in \citetads{2017A&A...606A.142C}, where data at 1.2 and 2 mm were available. This improvement is due to a better angular resolution and to the study of a wider wavelength range, which can better constrain the spectral index of the dust. 

The paper is divided as follows: in Section \ref{Observations} we describe the observations and data processing; in Section \ref{analysis} we present the two new continuum maps and show a first attempt of checking opacity and spectral index variations following the ratio between both bands and trying to model the emission seen using constant spectral index and opacity values towards the core; we also present a new model for the physical structure of L1544 using the observed opacity variations. We discuss these new findings in Section \ref{discussion}. We summarize our results in Section \ref{conclusions}.

\section{Observations} \label{Observations}

\subsection{AzTEC}

The data at 1.1 mm were obtained during the nights of February 22$^{\mathrm{nd}}$ and March 22$^{\mathrm{nd}}$ of 2016, at the Large Millimeter Telescope (LMT) Alfonso Serrano, placed at the volcano Sierra Negra, Mexico. The observations were carried out during the early science phase, when LMT had a diameter of 32\,m. The opacity at 220 GHz at zenith ranged from 0.06-0.08, and the total integration time was 1.6 hours. 
L1544 was observed with the continuum camera AzTEC \citepads{2008MNRAS.386..807W} and using Rastajous Map Mode \citepads{2018ApJ...852..106C}. The data reduction was done using the standard pipeline of AzTEC \citepads{2008MNRAS.385.2225S}, with the implementation of the Cottingham method \citepads{1987PhDT.........4C} to better recover the large scale structure of the core \citepads[see]{2018ApJ...852..106C}. The Cottingham method helps to remove the contamination from the atmosphere by modeling the temporal variations of the atmosphere signal using B-splines, and this is done for each detector. This method is a maximum likelihood estimator for the emission coming from the atmosphere and the astronomical sources \citepads{2010ApJS..191..423H}. 3C111 was observed during 120 seconds every hour for pointing calibration. The final map has a pointing error below 1\arcsec. CRL618 was used as beam map and flux calibrator, obtaining an absolute flux calibration uncertainty of 10\%. Studying the transfer function of the pipeline used for this set of observations, we are reliably recovering scales up to 3.6\arcmin\ in size (see Appendix \ref{transfer_function}). Compared to the 1.2 mm observations presented by \citetads{2017A&A...606A.142C}, this map improves the recoverable scales by a factor of 1.8. The resulting map has a resolution of 12.6\arcsec\ and an rms of 3 mJy/beam.

\subsection{MUSTANG-2}

L1544 was observed at 3.3 mm during the nights of the 3$^{\mathrm{rd}}$ and 7$^{\mathrm{th}}$ of January 2018, at the Green Bank Observatory (GBO; project GBT17B-174). The observations were carried out with the bolometer camera MUSTANG-2 \citepads{2014JLTP..176..808D,2016JLTP..184..460S}. The total on source integration time was 3.25 hours. The observations were done with the On-The-Fly (OTF) mapping mode \citepads{2007A&A...474..679M}, using a daisy-petal scan pattern. This method allows frequent crossings to subtract the atmosphere and other systematics. To accomplish this subtraction, a median common mode and slowly varying per-detector polynomial (30 seconds timescale) is subtracted from the data, excluding the central region of the map from the polynomial fit. Each night, at intervals of approximately 20 minutes, quick (94 second) observations of the nearby quasar J0510+1800 were made.  These were used to check the focus of the telescope, correct any pointing offsets, and to measure the MUSTANG-2 beam to high accuracy.  J0510+1800 is a flux calibrator for ALMA and was used as a check on the GBO's main beam efficiency.  Additional observations of J0750+1231 and J0754+2006 (also ALMA calibrators) were used as cross checks and gave consistent results.  We estimate an overall calibration error of 12\% (7\% due to source variability; 8\% due to beam volume variability; and 5\% due to the ALMA absolute flux calibration). The study of the transfer function for the pipeline used for the MUSTANG-2 data reduction shows that these observations recover scales up to $6.2\arcmin$ in size, which are modestly larger than the $4.2\arcmin$ instrumental instantaneous field of view (see Appendix \ref{transfer_function}). The final map has an rms of 0.15 mJy/beam, with a beam size of 9.7\arcsec. 

\subsection{Matching PSF}

The beam shape from MUSTANG-2 cannot be considered as Gaussian, in contrast with the beam of AzTEC. Therefore, in order to compare the two maps at the same resolution, we need to find a convolution kernel which transforms the PSF (Point Spread Function) of MUSTANG-2 to that of AzTEC. For this purpose, we made use of the Python library \textit{photutils.psf.matching}. We modified the shape of the window and the parameter for tapering (to remove the high frequency noise) until the convolved MUSTANG-2 beam matched the AzTEC PSF. 

\subsection{Herschel/SPIRE}

In Section \ref{modeling1}, the \textit{Herschel}/SPIRE maps of L1544 are used. They were presented by \citetads{2016A&A...592L..11S} and are part of the \textit{Herschel} Gould Belt Survey \citepads{2010A&A...518L.102A}. As these SPIRE 250, 350 and 500 $\mu$m maps are used together with ground-based telescope maps, they are artificially filtered in order to account for the missing flux coming from the large scale structure of the cloud. This method is based on subtracting background emission, and has been compared with more sophisticated methods for the case of L1544 in \citetads{2017A&A...606A.142C}, finding that the results of the analysis done with the two methods are in agreement within the errors. For more details on these maps and the filtering process, see \citetads{2017A&A...606A.142C}. We note that this only affects the results presented in Sect. \ref{modeling1}.

\section{Analysis and results}\label{analysis}\label{results}

\begin{figure*}
\includegraphics[width=1.0\textwidth]{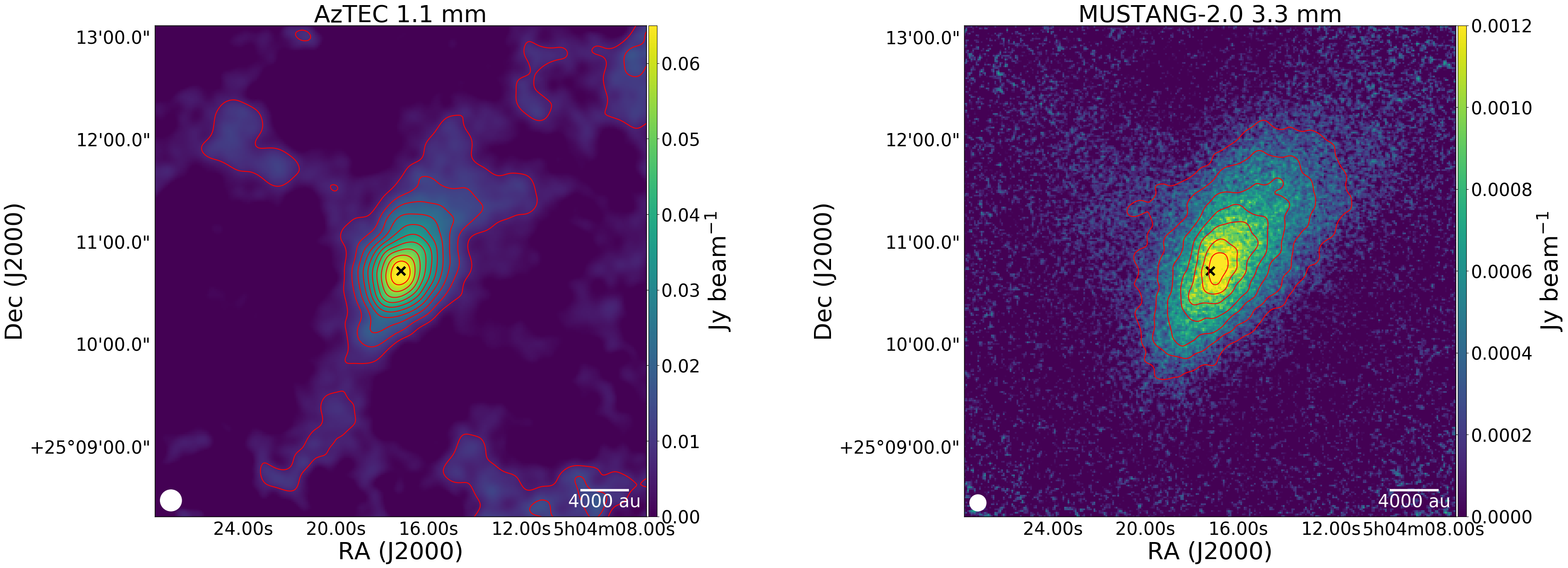}
\caption{\textit{Left panel:} AzTEC map at 1.1 mm. The contours represent steps of 2 $\sigma$, starting from 2 $\sigma$. The pixel size is 1\arcsec , the HPBW is shown in the bottom left corner, and the 1.3 mm dust continuum peak is shown with a black cross \citepads{1999MNRAS.305..143W}. \textit{Right panel:} MUSTANG-2 map at 3.3 mm. The contours represent steps of 2 $\sigma$, starting from 2 $\sigma$. The pixel size is 1\arcsec , the HPBW is shown in the bottom letf corner, and the 1.3 mm dust continuum peak is shown with a black cross \citepads{1999MNRAS.305..143W}. }\label{maps}
\end{figure*}

Fig. \ref{maps} shows the final maps, at their intrinsic resolution. This is the first time that ground-based continuum observations hint at the detection of a filamentary structure or extended emission towards the north-east in continuum, which are clearly seen by Herschel/SPIRE maps \citepads[see]{2016A&A...588A..30S} and C$^{18}$O map \citepads{1998ApJ...504..900T}. These maps are compared with those of NIKA \citepads{2017A&A...606A.142C}, for checking filtering effects, in Appendix \ref{comparison_nika}.

\subsection{Spectral index and opacity maps}\label{betakappa_map}
In pre-stellar cores the emission of the dust can be assumed to be optically thin at millimeter wavelengths, and it can be described by a modified black body function:

\begin{equation}\label{eqSS}
I_{\nu} =  \mu_{\mathrm{H_2}} m_{\mathrm{H}}   \int_{s_{los}}^{}{B_{\nu}[T_{d}(s)]n_{\mathrm{H_2}}(s)\kappa_{\nu}(s) \mathrm{d}s},
\end{equation} 
\\
where $I_{\nu}$ is the intensity, $B_{\nu}[T_{d}(s)]$ the blackbody function at a temperature $T_{d}(s)$, $\kappa_{\nu}(s)$ is the dust opacity at frequency $\nu$, $\mu_{\mathrm{H_2}}=2.8$ is the molecular weight per hydrogen molecule \citepads{2008A&A...487..993K}, $m_{\mathrm{H}}$ is the mass of the hydrogen atom, $n_{\mathrm{H_2}}(s)$ is the molecular hydrogen density, and $s$ is the line of sight depth. The dust opacity can be approximated by a power law at millimeter wavelengths \citepads{1983QJRAS..24..267H}:
$\kappa_{\nu} = \kappa_{\nu_0}\left( \frac{\nu}{\nu_{0}}\right)^{\beta}$,  where $\beta$ is the spectral index of the dust. Although laboratory studies have shown that this power law changes depending on the frequency range \citepads[e.g.]{2017A&A...606A..50D}, the study of the dependency of $\beta$ with frequency is beyond the scope of this work.  

If the density and the temperature profiles are known, the derivation of the spectral index and the opacity is straightforward with only data at two different frequencies. From Eq. \eqref{eqSS}, the spectral index can be derived from the ratio of the emission at both wavelengths, following:

\begin{equation}\label{beta_ratio}
\beta = \frac{\log (I_{1 \mathrm{mm}}/I_{3 \mathrm{mm}}) - \log (B_{1 \mathrm{mm}}[T_{\mathrm{d}}]/B_{3 \mathrm{mm}}[T_{\mathrm{d}}])}{\log (\nu_{1 \mathrm{mm}}/\nu_{3 \mathrm{mm}})},
\end{equation}
\\
where the sub-indexes  1 mm and 3 mm denote the band at which the intensity, black body emission and frequency are being evaluated. The opacity can be derived from any of the wavelengths observed using Eq. \eqref{eqSS}. A dust-to-gas ratio of 0.01 is assumed here. Although this ratio show variations between different regions of our galaxy \citepads{2011piim.book.....D}, we do not expect them to be significant within the same cloud.  

Our first analysis of the new data follows the methodology presented by \citetads{2017A&A...606A.142C}, where the temperature and density profiles from \citetads{2015MNRAS.446.3731K} were used for deriving the spectral index and the opacity following Eqs. \eqref{eqSS} and \eqref{beta_ratio}. \citetads{2015MNRAS.446.3731K} deduced the physical model of L1544 by following the evolution of an unstable Bonnor-Ebert sphere in quasi-equilibrium contraction and by comparing, via radiative transfer modeling, modeled and observed molecular line emission of low and high density tracers.

\begin{figure}
\includegraphics[width=0.5\textwidth]{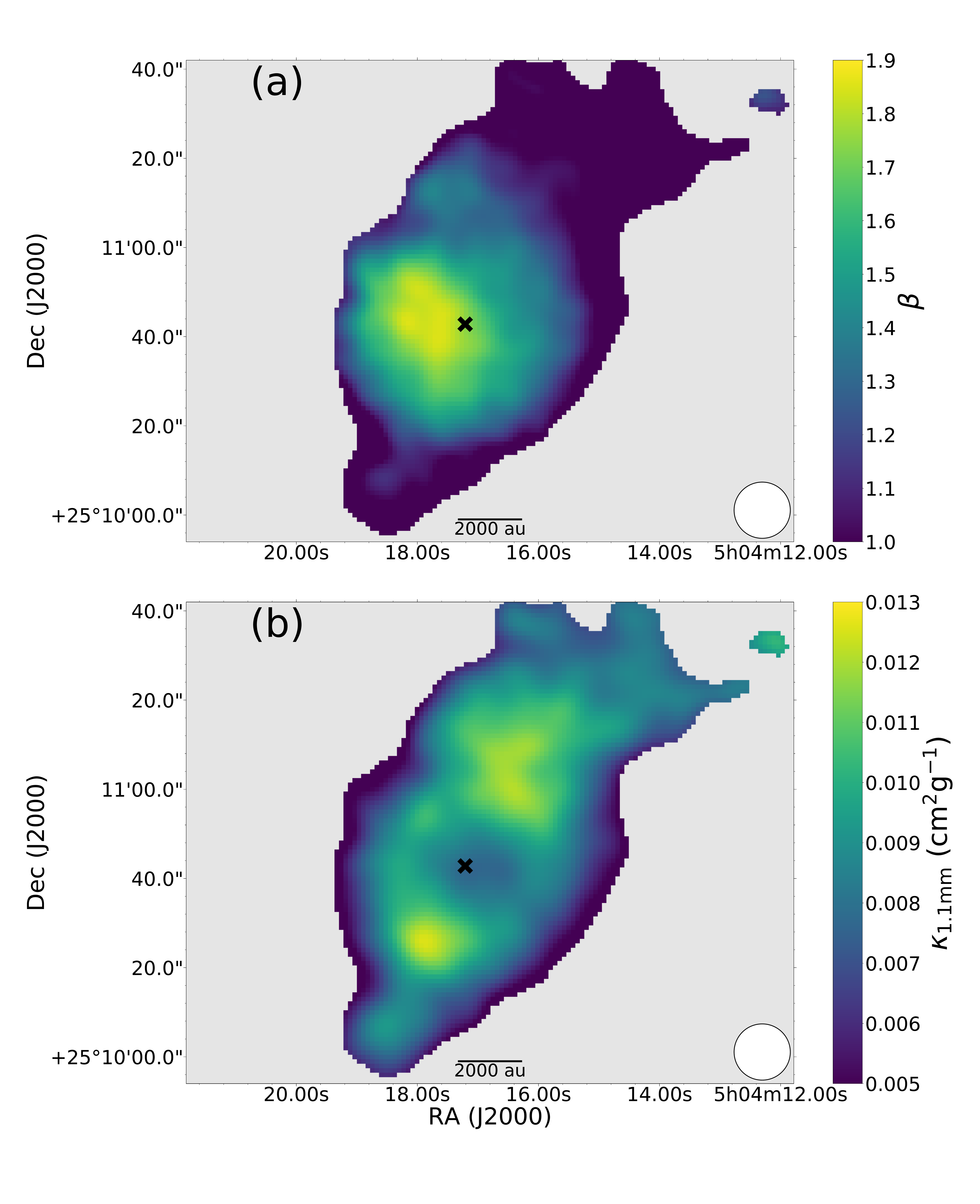}
\caption{\textit{Panel a)} shows the spectral index map, while \textit{panel b)} shows the dust opacity map. Both maps have been derived as described in Section \ref{betakappa_map}. The error in the spectral index is $\sim$0.2 and in the opacity is $\sim$10\%.}\label{map_beta_kappa}
\end{figure}

The resulting maps of the dust spectral index and opacity at 1.1 mm are shown in Fig. \ref{map_beta_kappa}. We find very similar results to those seen previously in \citetads{2017A&A...606A.142C}: while the spectral index increases towards the center, the opacity decreases. The emission at the two wavelengths bands peak at slightly different places, and this may cause the displacement of the spectral index peak from the 1.1 mm peak. The increase of the spectral index towards the center could be an indication of the presence of dust grains large enough to affect the emission at 1.1 mm, but not so much that it affects the emission at 3.3 mm. However, this would imply an increase in the opacity at 1.1 mm, and Fig. \ref {map_beta_kappa} shows a circular region around the center within which the opacity decreases. This is caused by the fact that the model is centrally concentrated. Moreover, as noted previously by \citetads{2017A&A...606A.142C}, the emission of the core is clearly not following a sphere (see Fig. \ref{maps}), and the model is spherical. \citetads{2017A&A...606A.142C} solved this problem by comparing the model with the emission of the core averaged in concentric ellipses. We therefore follow this same procedure in the following sections.

\subsection{New spectral index and opacity}\label{modeling1}

The emission of the core is averaged in concentric ellipses separated by 1.5\arcsec. The ratio of the major to minor axes of these ellipses is 1.5, and the major axis is inclined with respect to the declination by 65 degrees. The radial emission profiles are obtained by taking the geometric mean of the major and minor axes of the ellipses; the corresponding radius is labeled $r_m$. 

\begin{figure}
\includegraphics[width=0.5\textwidth]{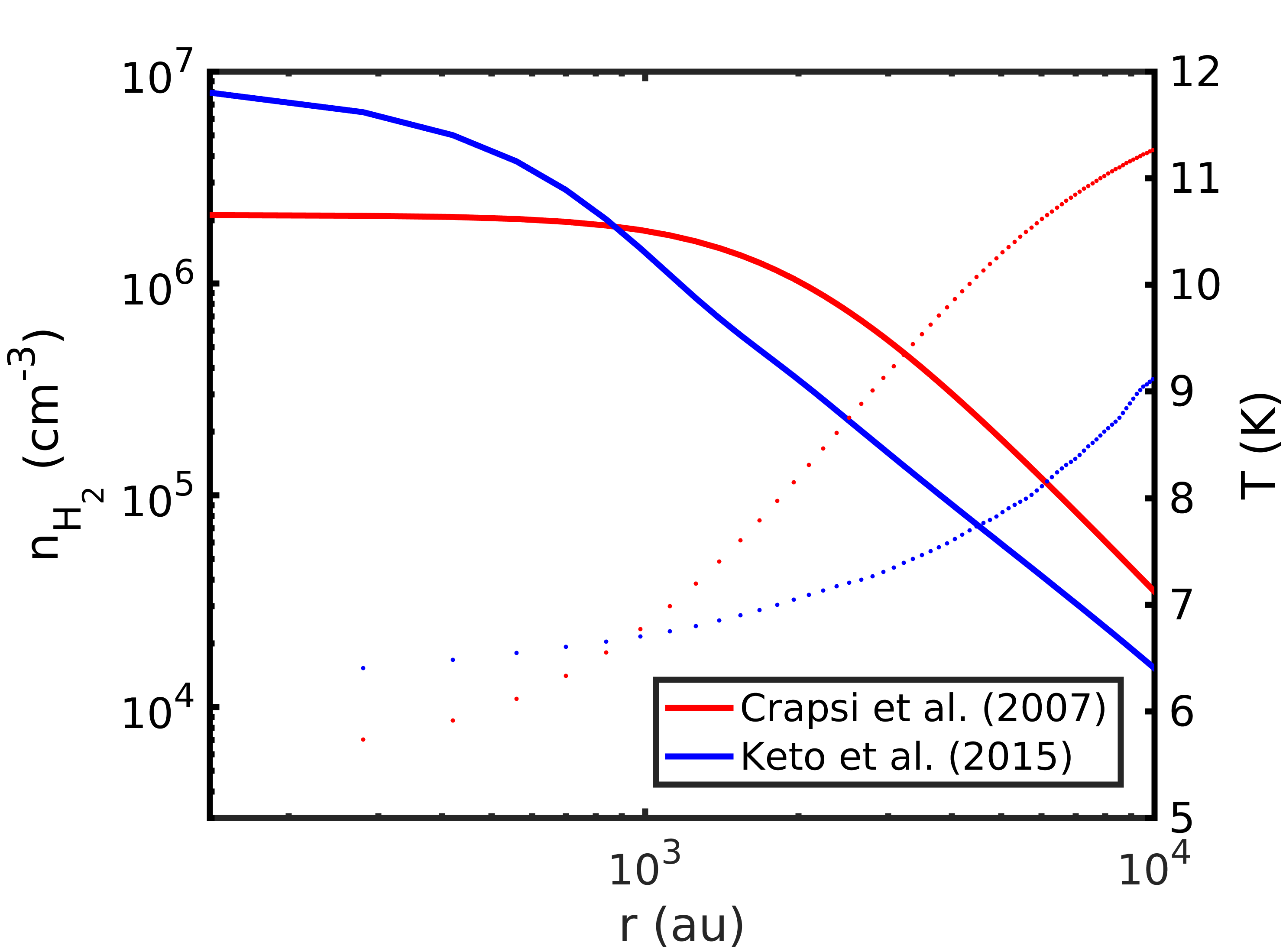}
\caption{Density (solid line) and temperature (dotted line) profiles describing the physical properties of L1544 by \citetads{2007A&A...470..221C} in red and by \citetads{2015MNRAS.446.3731K} in blue.}\label{models}
\end{figure}

We now include in our analysis the model for L1544 that \citetads{2007A&A...470..221C} presented, which is an improvement of the one presented by \citetads{2002ApJ...569..815T}. The temperature profile of this model was computed via excitation and radiative transfer modeling of their interferometric ammonia observations, while the density profile was constrained by fitting the emission seen in the 1.3 mm continuum map \citepads{1999MNRAS.305..143W}. Therefore, while \citetads{2015MNRAS.446.3731K} is purely theoretical, although constrained by observations,\citetads{2007A&A...470..221C} is based on observational results. Fig \ref{models} shows the differences between the two models.

Modeling the core emission first requires a choice of spectral index and opacity. Following \citetads{2017A&A...606A.142C}, we fit the spectral energy distribution (SED) of the core toward the center, using the emission obtained with AzTEC, MUSTANG-2 and Herschel/SPIRE, after smoothing and regridding all the data to the resolution of the 500$\mu$m band ($\sim$38.5\arcsec resolution maps with pixel size of 14\arcsec). The fit is performed towards the central pixel and is shown in the Appendix \ref{sed_fit_appendix}. This fit takes into account the temperature and density distributions in the core predicted by the physical models from \citetads{2007A&A...470..221C} and \citetads{2015MNRAS.446.3731K} as well as the color corrections for the SPIRE bands. For a detailed description of this procedure, see \citetads{2017A&A...606A.142C}. 

The resulting spectral indexes and opacities from the SED fits are: $\beta$ = 1.6 $\pm$ 0.4 and $\kappa_{250\mathrm{\mu m}}$ = 0.03 $\pm$ 0.01 cm$^2$g$^{-1}$ for the model of \citetads{2007A&A...470..221C}; and $\beta$ = 2.0 $\pm$ 0.4 and $\kappa_{250\mathrm{\mu m}}$ = 0.16 $\pm$ 0.07 cm$^2$g$^{-1}$ for the model of \citetads{2015MNRAS.446.3731K}. There is a difference of a factor of $\sim$5 between the opacities of the two models. This difference will be discussed at the end of this Section. Comparing these values with the ones from \citetads{2017A&A...606A.142C}, which are $\kappa_{250\mathrm{\mu m}} = 0.2 \pm 0.1$ cm$^2$g$^{-1}$ and $\beta=2.3 \pm 0.4$, we find that the spectral index is lower, although consistent within the errors. The slight difference is caused by the different filtering process applied during the data reduction process to the millimeter maps.  In \citetads{2017A&A...606A.142C}, the NIKA maps at 1.2 and 2 mm suffered from substantial filtering, which implied that, when smoothing the data to bigger beams, the emission was reduced due to the inclusion of negative flux values in the dust peak. Moreover, \citetads{2017A&A...606A.142C} expected to be recovering the emission from spatial scales smaller than 2\arcmin; while in these new maps we estimate to be recovering the emission from spatial scales up to 3.6\arcmin. This produces a higher spectral index. 

\begin{figure}
\includegraphics[width=0.5\textwidth]{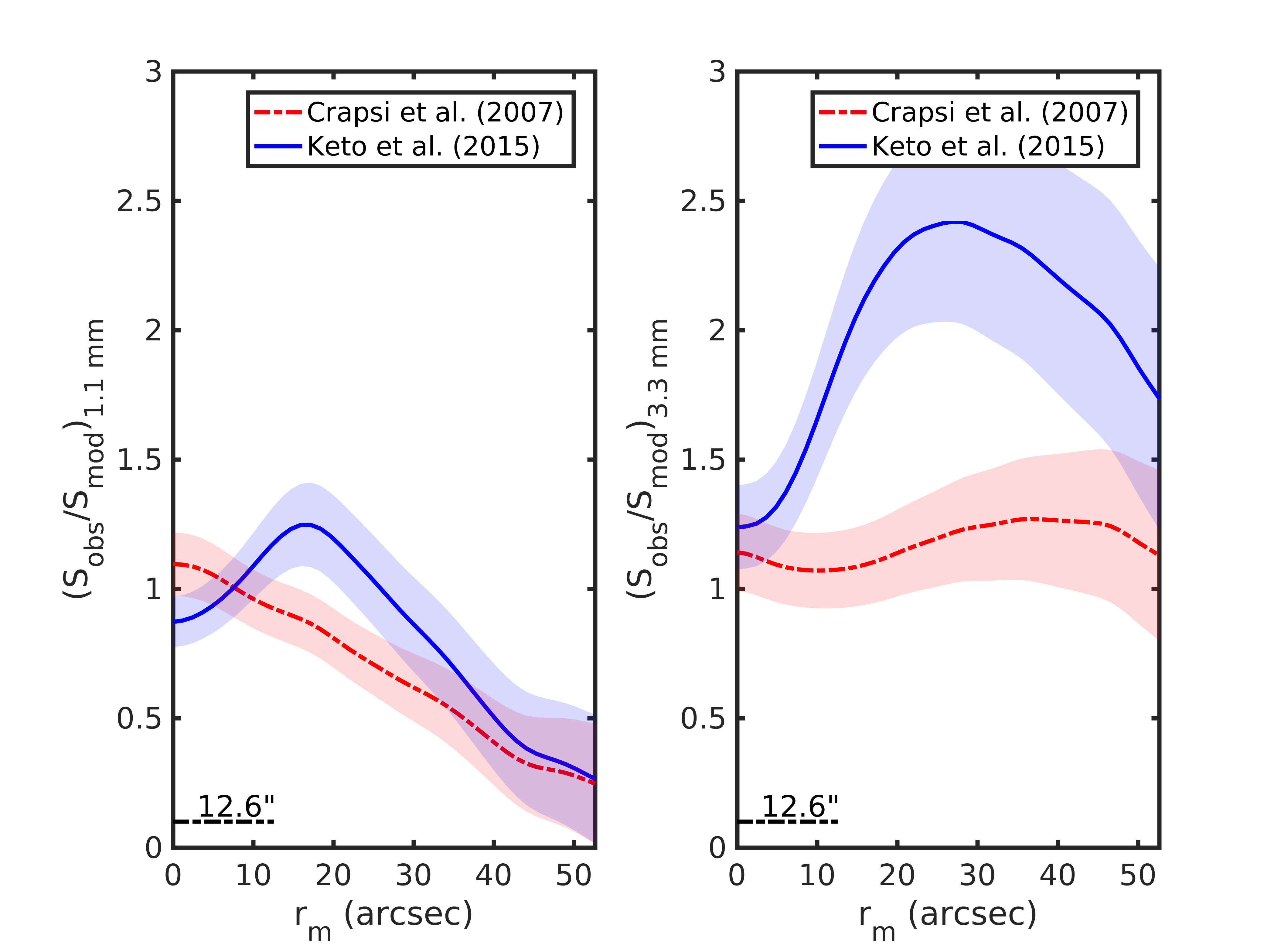}
\caption{Ratio between the observed emission and the modeled emission derived as described in Section \ref{modeling1}, as a function of projected radius $r_m$. The shaded regions show the error associated with the data. The resolution of the maps is indicated with a bar of length 12.6\arcsec in the bottom left corner of both panels. }\label{ratio}
\end{figure}

With these new spectral indexes and opacities we proceed to check whether the models can reproduce the observations. We adopt a constant spectral index and opacity. Fig. \ref{ratio} shows the ratio between the observations and the modeled emission. The first thing to notice is that the model of \citetads{2015MNRAS.446.3731K} does not reproduce the data, showing discrepancies between the model and the observations of up to a factor of 2 for the 1.1 mm band and a factor of 2.5 in the 3.3 mm band within the inner 36\arcsec (region where the emission of the core is detected above 3 $\sigma$).  Changing the absolute values of the opacity and spectral index, for example, using the ones from \citetads{2017A&A...606A.142C}, does not solve the situation (see Appendix \ref{modeling_appendix}). The comparison between the model of \citetads{2007A&A...470..221C} and the observations shows that the 1.1 mm band is badly reproduced, with differences of a factor of 2 in the outer parts of the core; while the model can reproduce the emission at 3.3 mm within 20\%. 
These results therefore indicate that either the models are wrong or that our assumption of constant spectral index and opacity across the cloud is not valid.

If the models are correct, then the required variations in the opacity, when considered constant along the line of sight, can be obtained from Eq. \eqref{eqSS}. This opacity can be considered the averaged opacity along the line of sight ($\overline{\kappa}_{\nu}$), as it is the opacity weighted by the temperature and density. Then, $\overline{\kappa}_{\nu}$ can be derived following the ratio for both wavelengths:
\begin{equation}
\overline{\kappa}_{\nu} =  \frac{I_{\nu}}{\mu_{\mathrm{H_2}} m_{\mathrm{H}}   \int_{s_{\mathrm{los}}}^{}{B_{\nu}[T_{\mathrm{d}}(s)]n_{\mathrm{H_2}}(s) ds}}.
\end{equation} 

\begin{figure}
\includegraphics[width=0.5\textwidth]{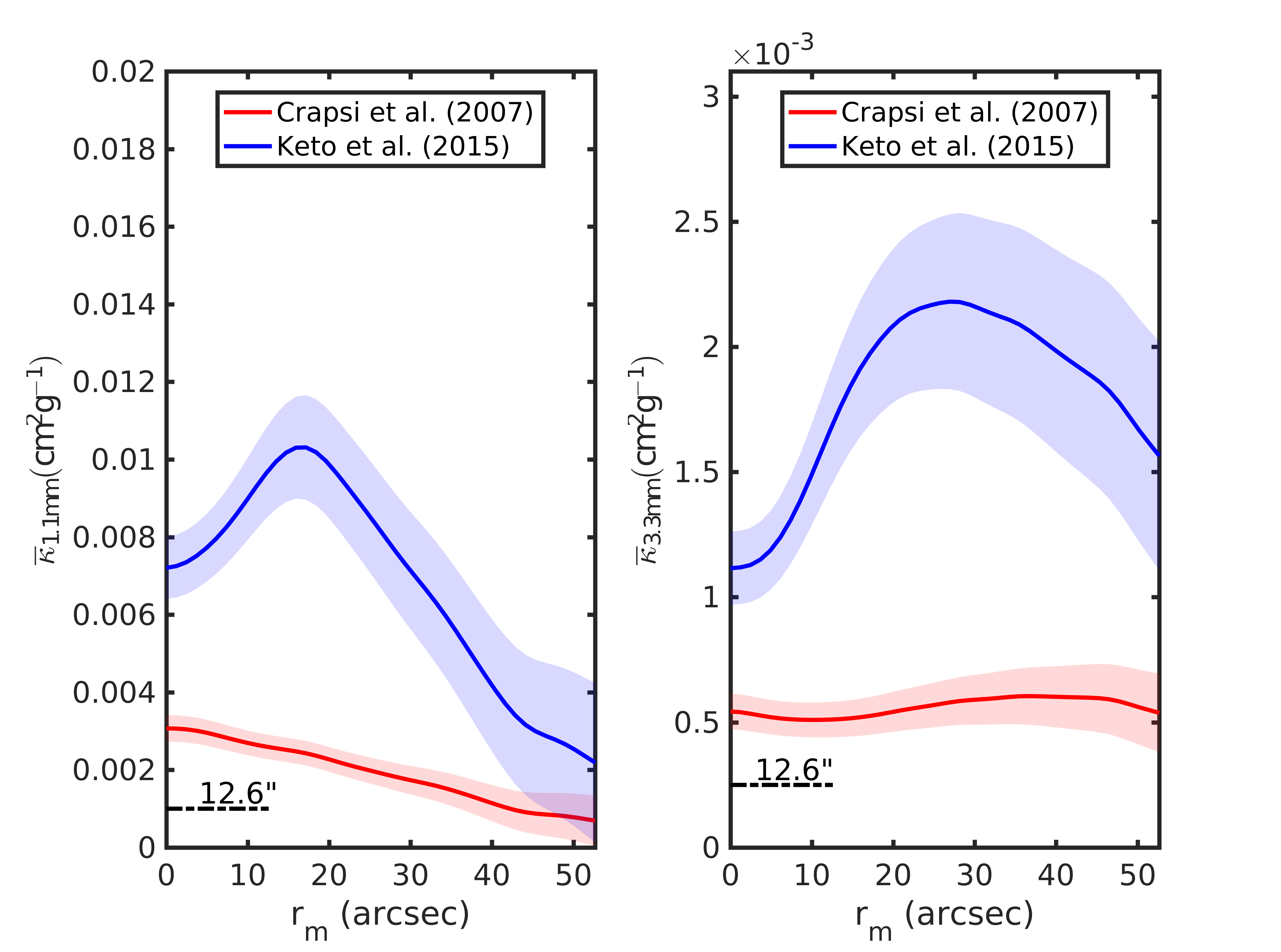}
\caption{Opacities at 1.1mm (left) and at 3.3 mm (right), averaged along the line of sight, as a function of the projected radius $r_m$. The blue curve is obtained when using the core physical structure derived by \citetads{2015MNRAS.446.3731K}; the red curve is obtained when using the physical structure from \citetads{2007A&A...470..221C}. The shaded regions show the error associated with the data. The resolution of the maps is indicated with a bar of length 12.6\arcsec in the bottom left corner of both panels. }\label{opacity_variations}
\end{figure}

\begin{figure}
\includegraphics[width=0.5\textwidth]{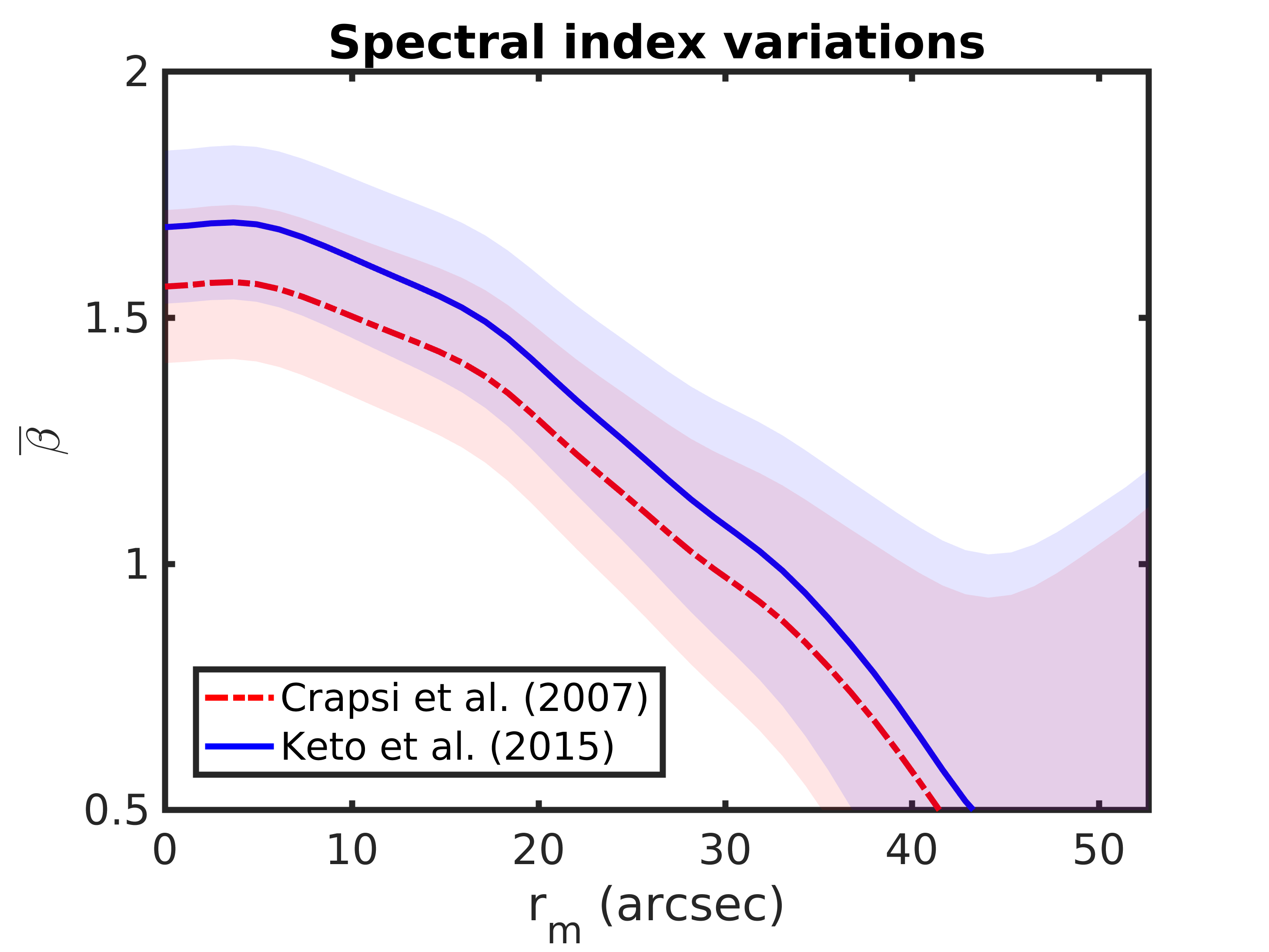}
\caption{Spectral index variation, averaged along the line of sight, as a function of the projected radius $r_m$, caused by the variation on $\overline{\kappa}_{\nu}$ shown in Fig. \ref{opacity_variations}. The different colors refer to the different physical structure adopted (blue for \citeads{2015MNRAS.446.3731K} and red for \citeads{2007A&A...470..221C}). The shaded regions show the error associated with the data. }\label{spectral_index_variations}
\end{figure}

Fig. \ref{opacity_variations} shows the variations of the opacities. The gradients in $\overline{\kappa}_{\nu}$  imply variations in the spectral index, averaged along the line of sight (see Fig.  \ref{spectral_index_variations}). Figs. \ref{opacity_variations} and \ref{spectral_index_variations} show that there are substantial changes in the opacity and the spectral index averaged along the line of sight. 

To obtain radial variations of $\kappa_{\nu}$ and $\beta$, we follow the method described in \citetads{2014A&A...562A.138R}. Using the Abel transform \citepads[see e.g.]{1986ftia.book.....B} we can write Eq. \eqref{eqSS} in the following way: 

\begin{equation}\label{abel}
 \mu_{\mathrm{H_2}} m_{\mathrm{H}} B_{\nu}[T_{d}(r)]n_{\mathrm{H_2}}(r)\kappa_{\nu}(r) = -\pi^{-1} \int_{r}^{\infty} \dfrac{\mathrm{d}I_{\nu}}{\mathrm{d}b} \dfrac{\mathrm{d}b}{\sqrt{b^2-r^2}}  ,
\end{equation}
where $b$ is the projected distance to the center. This way, we are able to obtain opacity radial profiles once the temperature and the density are well defined. 


The procedure is very sensitive to noise, so we fit the emission profiles with an analytic function. In this manner, abrupt changes in the derivatives of the emission profiles are avoided. We used a combination of 3 Gaussian functions, which provides continuous derivatives and good fits to the emission profiles.  We note that this process gives radial profiles smoothed to the resolution of the data \citepads{2014A&A...562A.138R}, so the temperature and density profiles are smoothed to the resolution of 12.6\arcsec for consistency. To evaluate the error associated with this process, we first create 1000 maps for each wavelength, which are the result of adding in each map random noise from a Gaussian distribution of $\sigma$ equal to the rms of the data. Then, we derive the opacity variations for all the maps and assume as error the standard deviation of all the samples. 

\begin{figure}
\includegraphics[width=0.5\textwidth]{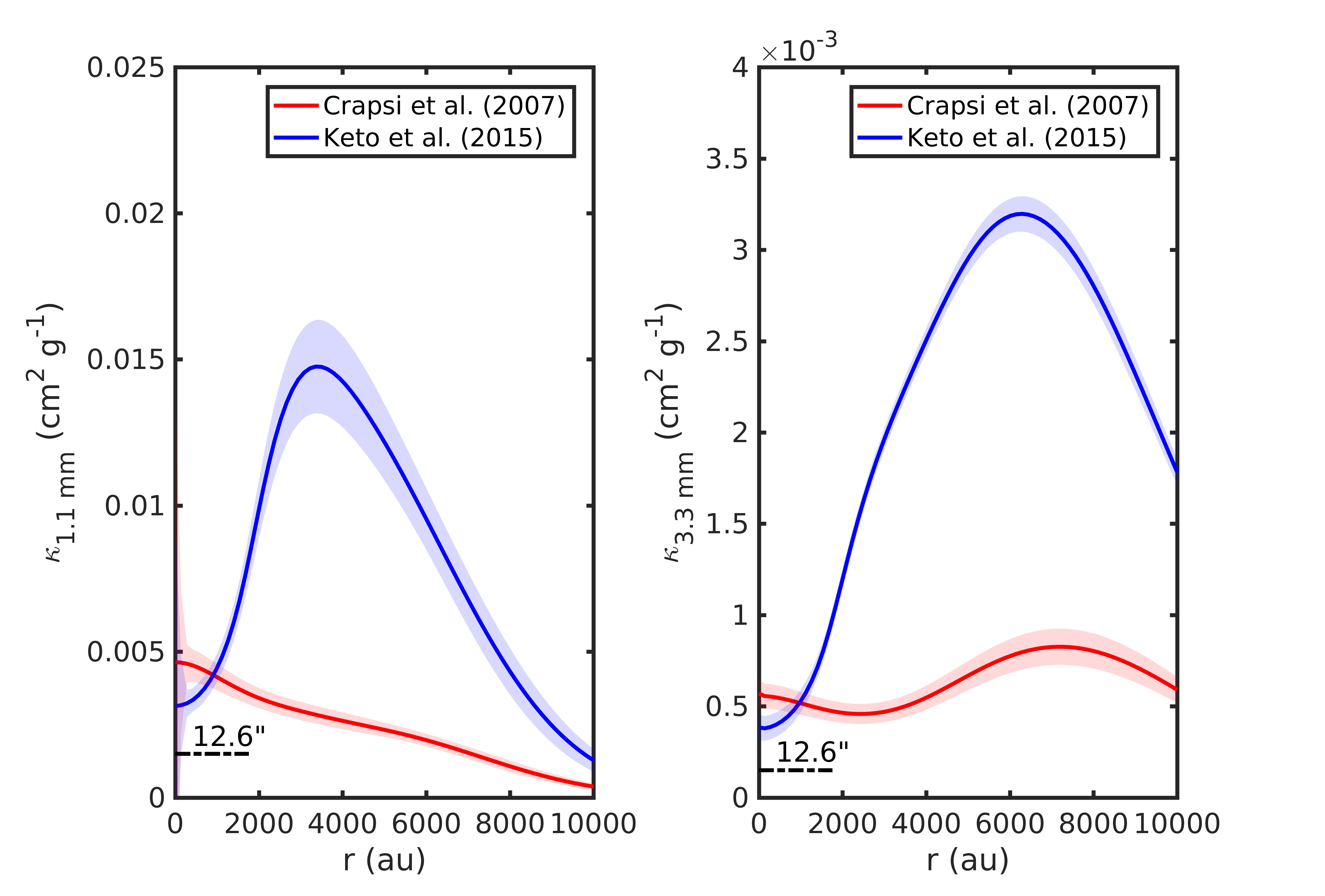}
\caption{Opacity radial variations obtained as explained in Section \ref{modeling1}. This figure shows $\kappa_{\nu}(r)$, while Fig. \ref{opacity_variations} shows $\kappa_{\nu}(r)$ averaged along the line of sight, i.e.,  $\overline{\kappa}_{\nu}(r_m)$. The different colors refer to the different physical structure adopted (blue for \citeads{2015MNRAS.446.3731K} and red for \citeads{2007A&A...470..221C}). The shaded regions show the error associated with the process. }\label{opacity_r}
\end{figure}

\begin{figure}
\includegraphics[width=0.5\textwidth]{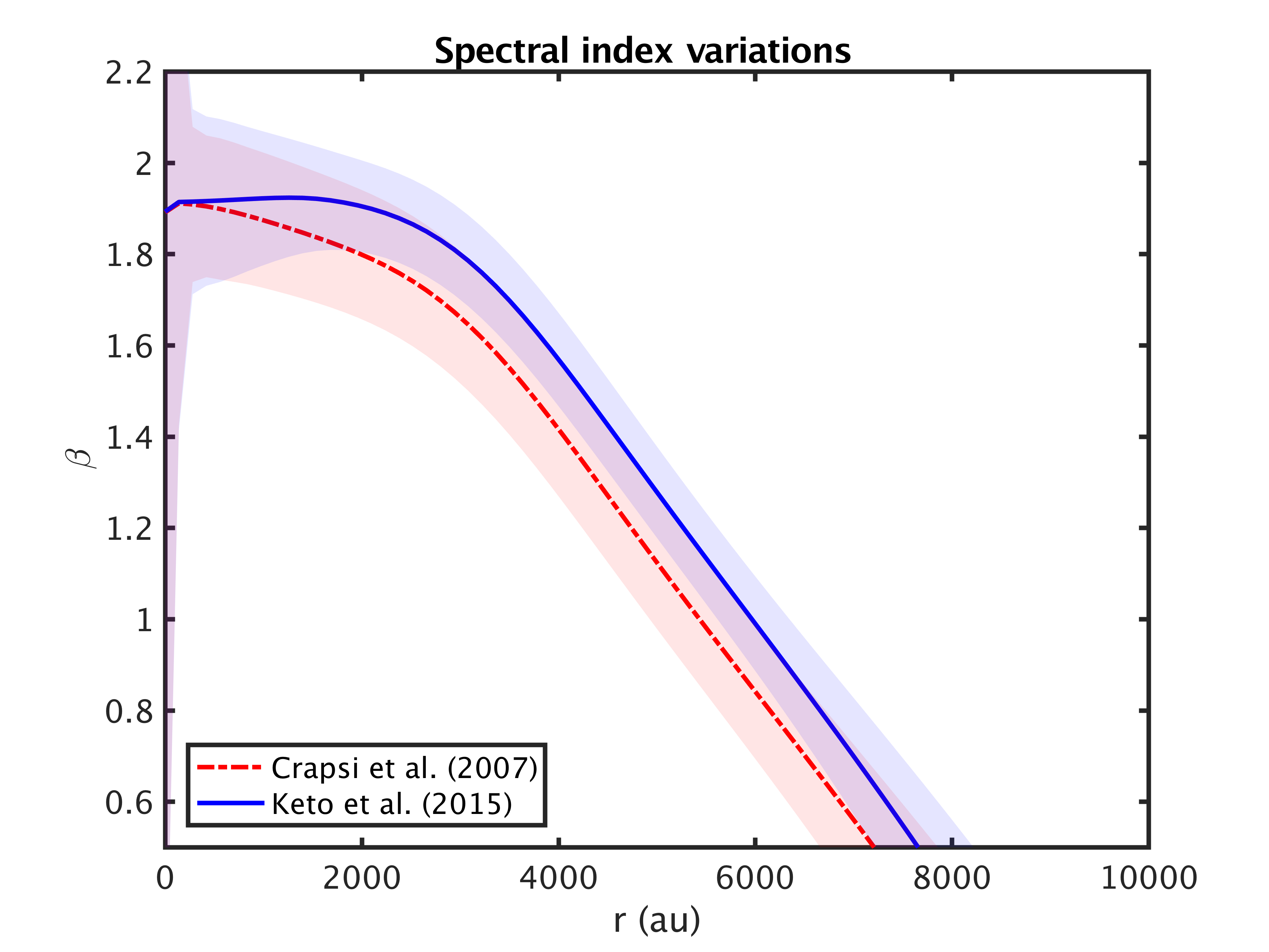}
\caption{Spectral index radial variations obtained as explained in Section \ref{modeling1}. This figure shows $\beta(r)$, while Fig. \ref{spectral_index_variations} shows $\beta(r)$ averaged along the line of sight, i.e., $\overline{\beta}(r_m)$. The different colors refer to the different physical structure adopted (blue for \citeads{2015MNRAS.446.3731K} and red for \citeads{2007A&A...470..221C}). The shaded regions show the error associated with the process. }\label{beta_r}
\end{figure}

Figs. \ref{opacity_r} and \ref{beta_r} show the resulting radial distributions of the opacities and the spectral index, respectively. The 1.1 mm opacities are in the range of values predicted by \citetads{1994A&A...291..943O} for different grain size distributions and conditions. The radial opacity profile for the model of \citetads{2015MNRAS.446.3731K}, which is consistent with dense clouds and thick ice mantels, follows a shape that indicates that the model produces too much emission in the center. On the other hand, the opacities at 1.1 mm for the model of \citetads{2007A&A...470..221C} are consistent with bare grains and no coagulation, although one has to take into account that a factor of 2 is within the uncertainties \citepads{1994A&A...291..943O}. The shape of the opacity at 3.3 mm is due to noise, although the extended emission towards the north-east, which is not taken into account in the model, tends to increase the opacity at large radii. Better sensitivity observations should help to improve on this. The spectral indexes show very similar behavior, which indicates that they mainly depend on the relative variation of the emission seen between both wavelengths. 

\begin{figure}
\includegraphics[width=0.5\textwidth]{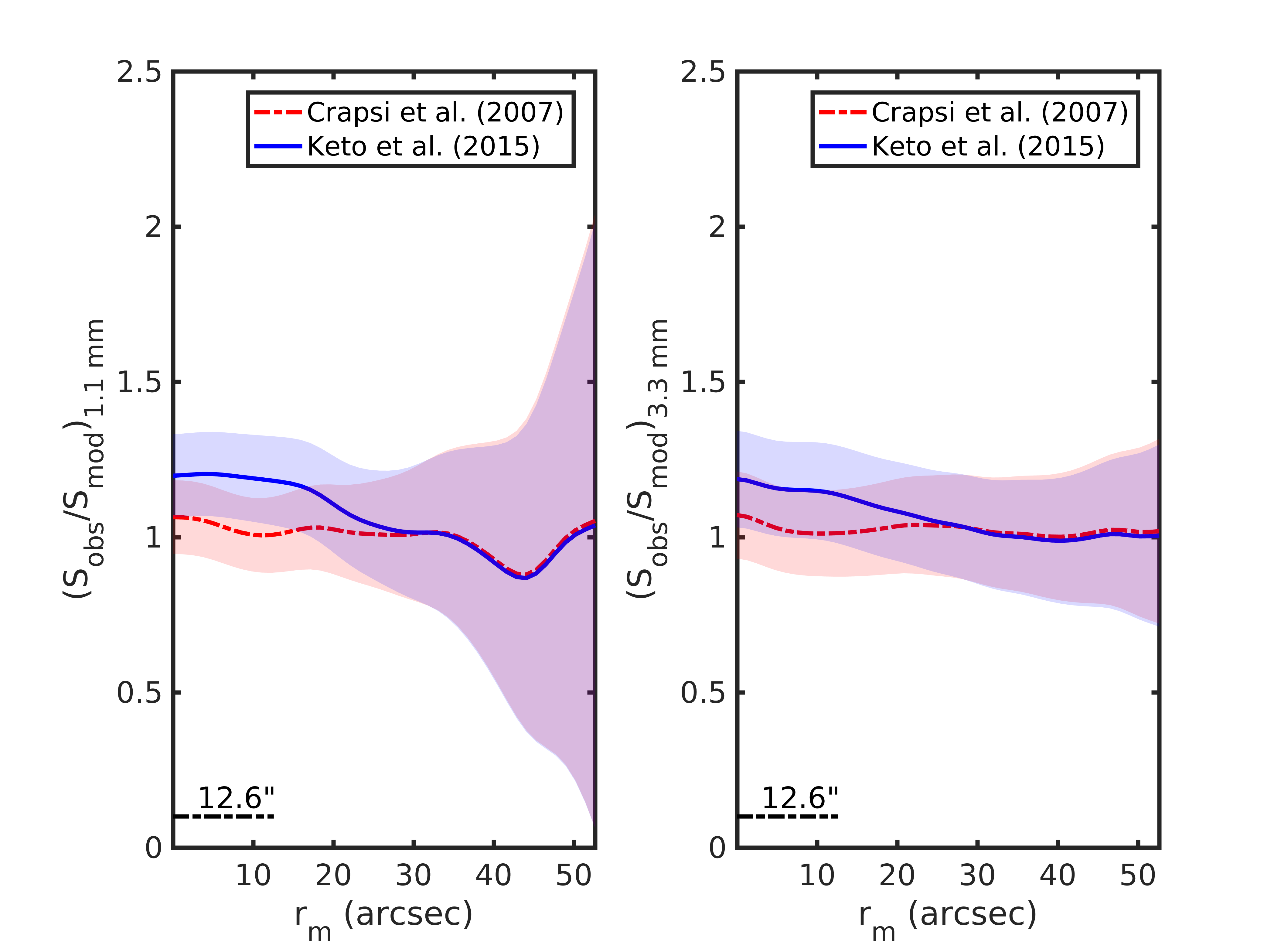}
\caption{Ratio between the observed emission profiles and the modeled emission profiles taking into account the radial opacity variations shown in Fig. \ref{opacity_r}. The shaded regions show the noise associated with the data.}\label{mm-kr}
\end{figure}

To examine the validity of the derived opacities, we generate synthetic maps and compare their emission profiles with the observed ones. The models reproduce the observations fairly well (see Fig. \ref{mm-kr}). In this process, the resolution of the models and the opacities were considered to be the same. Although the resolution of our observations does not allow us to resolve the inner 1\,000 au, where the difference between the two models is higher, the emission produced by them at a resolution of 12.6\arcsec is very different (as seen in Fig. \ref{ratio}), with the \citetads{2007A&A...470..221C} profile matching the data better than the \citetads{2015MNRAS.446.3731K} model.  The \citetads{2007A&A...470..221C} model was derived from observations so it has also a limited resolution of 7\arcsec, which is close to the resolution of the maps presented. Nevertheless, a discrepancy of 10-20\% is reasonable due to beam effects \citepads{2014A&A...562A.138R}. 


We underscore here that the opacity and spectral index variations found in this analysis depend on the particular density and temperature profiles assumed and inaccurate profiles will provide inaccurate and artificial variations in $\kappa_{\nu}$ and $\beta$. For example, as already said, \citetads{2015MNRAS.446.3731K} artificially increased the dust opacity by a factor of 4 to reproduce the temperature drop measured by  \citetads{2007A&A...470..221C}, as their model is dynamic and the gravitational compression toward the center produced extra heating; this dust opacity enhancement was applied throughout the core, affecting the overall structure. It is therefore natural that their physical model results, on average, in higher opacities than the model of \citetads{2007A&A...470..221C}. This shows that the initial assumption done for the opacity biases our results. In addition, opacity variations would indicate that any previous derivation of the temperature and density of the cloud which did not take this into account might also need modifications. In what follows, we use the model from \citetads{2007A&A...470..221C}, which does not consider dynamics and only aims at reproducing the observed temperature structure, as the starting point in our efforts to determine the radial distributions of the density, temperature, and the dust opacity in L1544.

\subsection{New physical structure}\label{new_method_model}\label{new}

For obtaining density, temperature and opacity profiles consistently, we apply an iterative method. Starting from the radial opacity variations shown in Fig. \ref{opacity_r} and the density distribution derived by \citetads{2007A&A...470..221C}, the following steps are done:

\begin{enumerate}
\item First, the optical depth is derived taking the radial profile of the opacity and the density\footnote{The optical depth at 1.1 mm, $\tau_{1 mm}$, always satisfies $\tau_{1 mm} \ll 1$, justifying the assumption of optical thinness throughout this study. }. From the optical depth we follow the equations given by \citetads{2001A&A...376..650Z} for deriving $A_V(r)$ and $T(r)$. These authors derived analytical equations that can be used to obtain the temperature profile for clouds externally illuminated by the standard interstellar radiation field (which includes an optical and near infra-red component coming from the emission of disk dwarf and giant stars, a far infra-red component from dust grains, a mid-infrared component from non-thermally heated grains and a millimeter component from the cosmic background radiation, following the work of \citeads{1994ASPC...58..355B} and \citeads{1983A&A...128..212M}). 
\item With this new temperature profile and the opacity variations  we derived a new density profile using the Abel transform (see Eq. \eqref{abel}). 
\item A new opacity profile is obtained from the new temperature and density profiles, again using the Abel transform. 
\end{enumerate}

From point 3, we return to point 1 until $\kappa_{1\mathrm{mm}}$ varies less than 0.001\%. In points 1 and 2 we need to take one band as reference, which we choose to be the one at 1.1 mm. However, these results are independent of the chosen band. In point 1 an extra factor in $A_V$ is included, which comes from the fact that the cloud is surrounded by an external layer of low density material which increments the value of $A_V$ by 2 magnitudes (see Appendix \ref{av_herschel}). Unfortunately, \citetads{2001A&A...376..650Z} do not provide a value for the temperature at extinctions lower than 10 magnitudes. Thus, we assume that the temperature follows a similar parametrization than that used by  \citetads{2007A&A...470..221C} and used their external temperature as a constraint in our temperature profile. Therefore, the temperature follows this formula:
\begin{equation}
T(r) = T_{out} - \frac{T_{out}-T_{in}}{1+\left(\frac{r}{r_{t0}}\right)^{\alpha_t}},
\end{equation}
\\
with $T_{out}$=12\,K, the temperature of the outer part of the core as measured by \citetads{2007A&A...470..221C}. We thus fix this value and fit the rest of the parameters to our data in each iteration.

Beam effects are not considered here. However, as already mentioned, the resulting profiles from this process will be smoothed with the beam of our observations. 

Only a few iterations are needed to find a convergence. The density profile is parametrized in the following way:

\begin{equation}
n_{\mathrm{H}_2}(r) = \frac{n_0}{1+\left(\frac{r}{r_0}\right)^{\alpha}}.
\end{equation} 

The obtained temperature and density profiles are:

\begin{equation}
T(r) = 12 (\mathrm{K}) - \frac{12 (\mathrm{K}) - 6.9 (\mathrm{K})}{1+\left( \frac{r(\mathrm{\arcsec})}{28.07\arcsec} \right)^{1.7}}
\end{equation}
and 
\begin{equation}
n_{\mathrm{H}_2}(r) = \frac{1.6\times10^{6} (\mathrm{cm^{-3}})}{1+\left(\frac{r(\mathrm{\arcsec})}{17.3 \arcsec}\right)^{2.6}}
\end{equation}

We note that this method is biased by the initial parameters used. A lower initial density leads to higher opacities, and vice versa. This issue is discussed in Appendix \ref{density_option}. The emission at only two wavelengths is fitted with three parameters, the density, temperature and dust opacity, and therefore, the solution is not unique. Therefore, further modeling efforts, utilizing all the available continuum data are needed.

\begin{figure}
\includegraphics[width=0.5\textwidth]{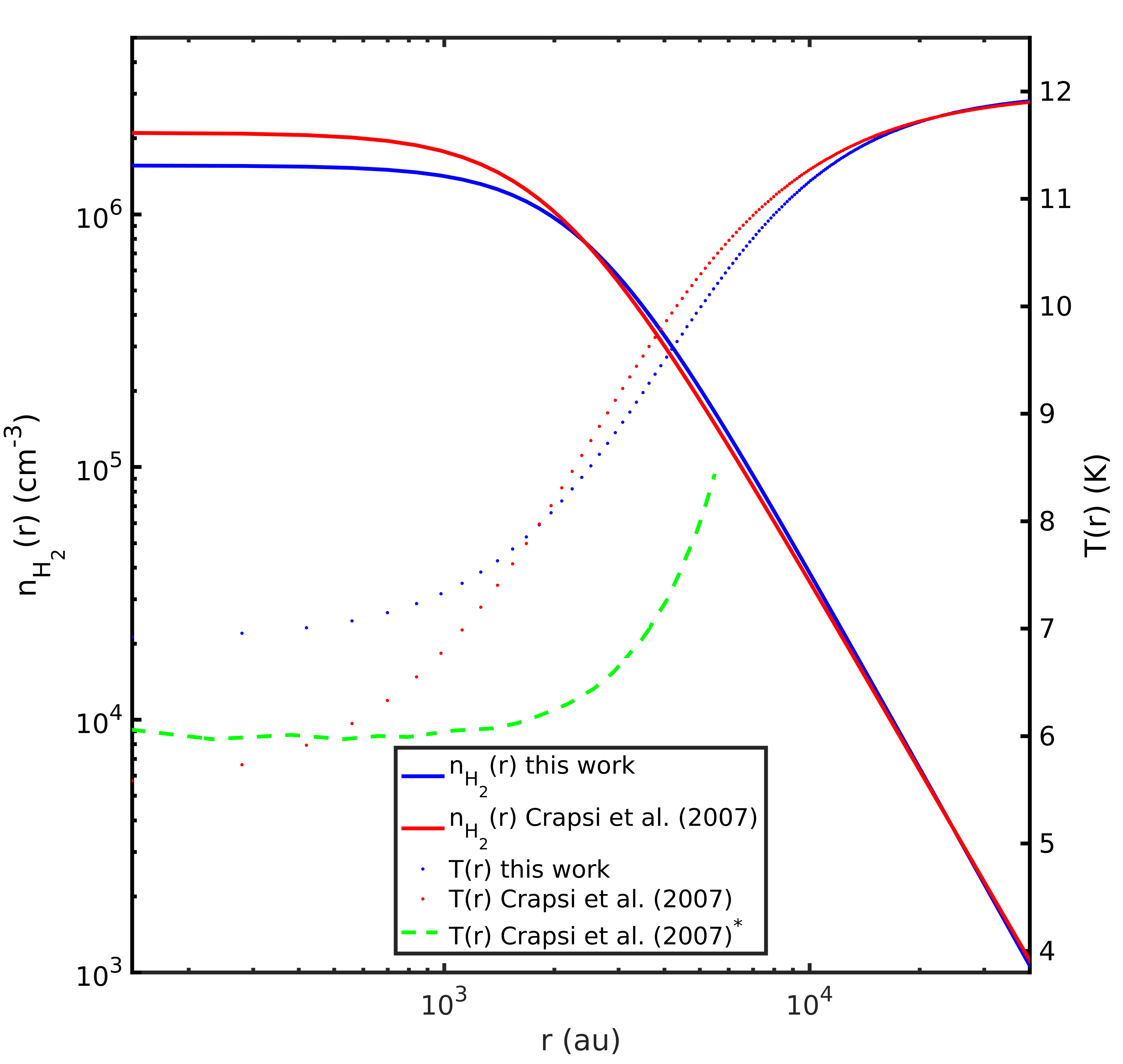}
\caption{New density (blue solid line) and temperature (blue dotted line) profiles derived as described in Section \ref{new}. In red, density (solid line) and temperature (dotted line) profiles from \citetads{2007A&A...470..221C}. In green, temperature profile from \citetads{2007A&A...470..221C} derived using the formulae from \citetads{2001A&A...376..650Z} (their Fig. 4), here \citetads{2007A&A...470..221C}$^{\ast}$. }\label{new_profiles}
\end{figure}

Fig. \ref{new_profiles} shows these new profiles, together with the profiles obtained by \citetads{2007A&A...470..221C} and the temperature profile obtained by \citetads{2007A&A...470..221C} using the method from \citetads{2001A&A...376..650Z} for comparison. 
On the one hand, there is a difference between the temperature derived by \citetads{2007A&A...470..221C} and the new temperature. This is because they obtained the temperature profile from gas temperature measurements, while here it is purely dust temperature. If our temperature profile is compared to their dust temperature profile, which was derived in a similar way to ours, there is better agreement. The difference is less than 1\,K, and it is due to the lower values in the opacities used here (compared to those commonly used), which result in lower values of $A_V$ towards the central regions (with a maximum of 52 magnitudes). The fact that the iteration did not lead us very far from the physical model of \citetads{2007A&A...470..221C} makes the solution plausible. 

\begin{figure}
\includegraphics[width=0.5\textwidth]{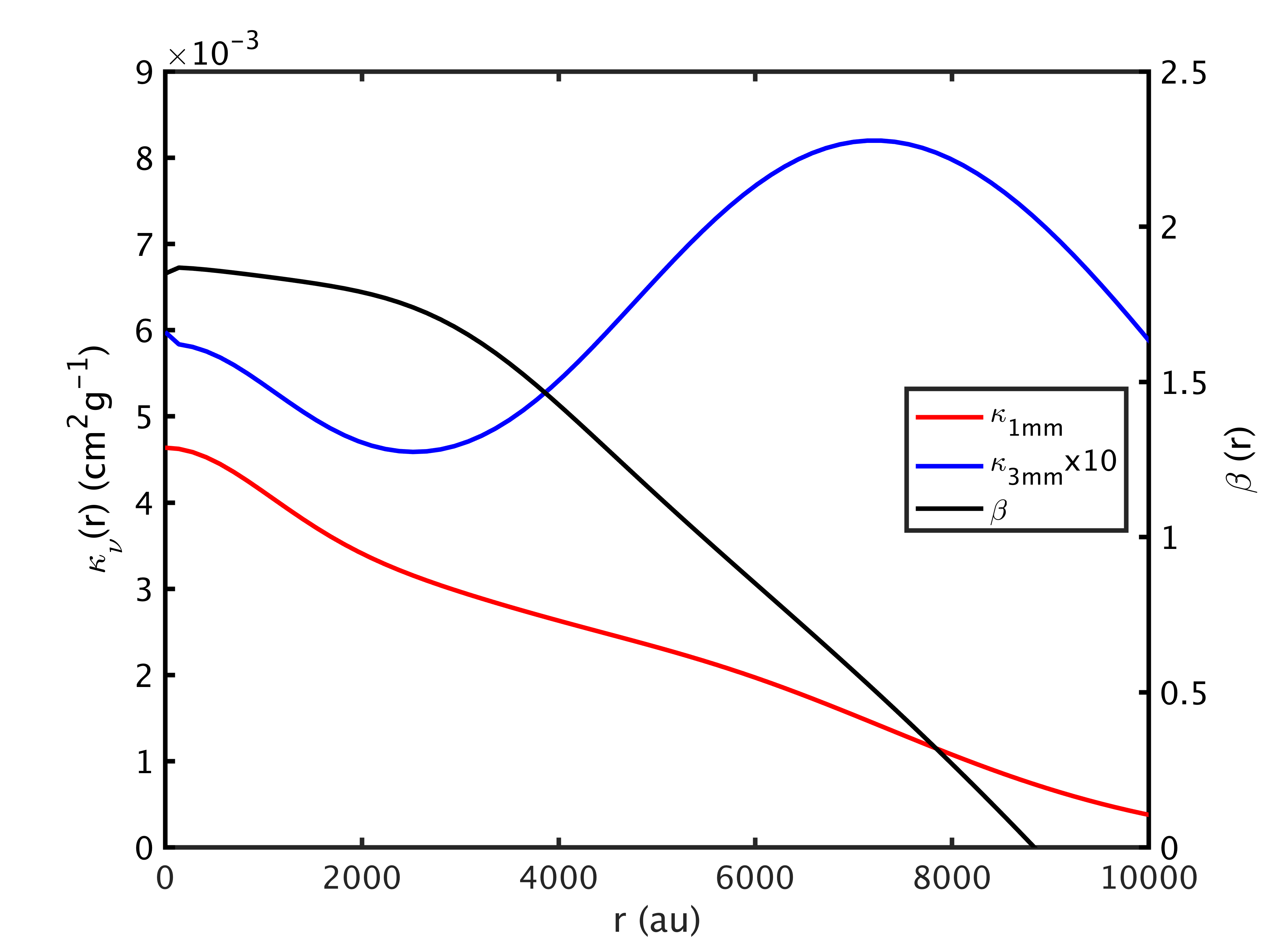}
\caption{Radial opacity and spectral index variations derived as described in Section \ref{new}.  }\label{kb-new}
\end{figure}

Fig. \ref{kb-new} shows the obtained radial distributions of the opacity and spectral index. The first thing to note is that the opacities go to 0 at large radii because of the method used. The Abel transform forces the left hand part of Eq. \eqref{abel} to be zero when the derivative of the emission with respect the impact parameter is 0. This is satisfied as soon as the emission is below 1$\sigma$. As $n_{\mathrm{H_2}}$ and $B[T_{\mathrm{d}}(r)]$ cannot be 0 due to their parametrization, the only parameter which can go to 0 is the opacity. Nevertheless, this is artificial, and we know that the cloud still emits at larger scales thanks to \textit{Herschel} observations. However, although the filtering does not seem to be a problem in these maps, the extended emission is very difficult to recover with ground based telescopes (we estimate to be recovering the emission of scales up to $\sim$3.6\arcmin at 1.1 mm), and this effect should be taken into account (by, for example, applying the same data reduction process to the synthesized maps). In any case, the variations derived for $r < 5\,000$ au, where the emission is above the 3 $\sigma$ level, are significant. This has been tested against the opacity and spectral index obtained if we filter the 3.3 mm map with the 1.1 mm map in the Fourier plane, in a similar manner as done in \citetads{2016A&A...588A..30S} and in the Appendix A of \citetads{2017A&A...606A.142C}, in order to check whether the different field of views could modify our results. The results using these filtered 3.3 mm map remain consistent within the uncertainties for $r < 5\,000$ au.

We derive the opacity at 3.3 mm also at its intrinsic resolution. This can be done because for deriving the opacity both wavelengths are treated independently, and the resolution of the map at 3.3 mm is only 23\% lower than that of the map at 1.1 mm, otherwise the density and temperature should change. Fig. \ref{kappa_3mm_intrinsic} shows the resulting opacity. 

\begin{figure}
\includegraphics[width=0.5\textwidth]{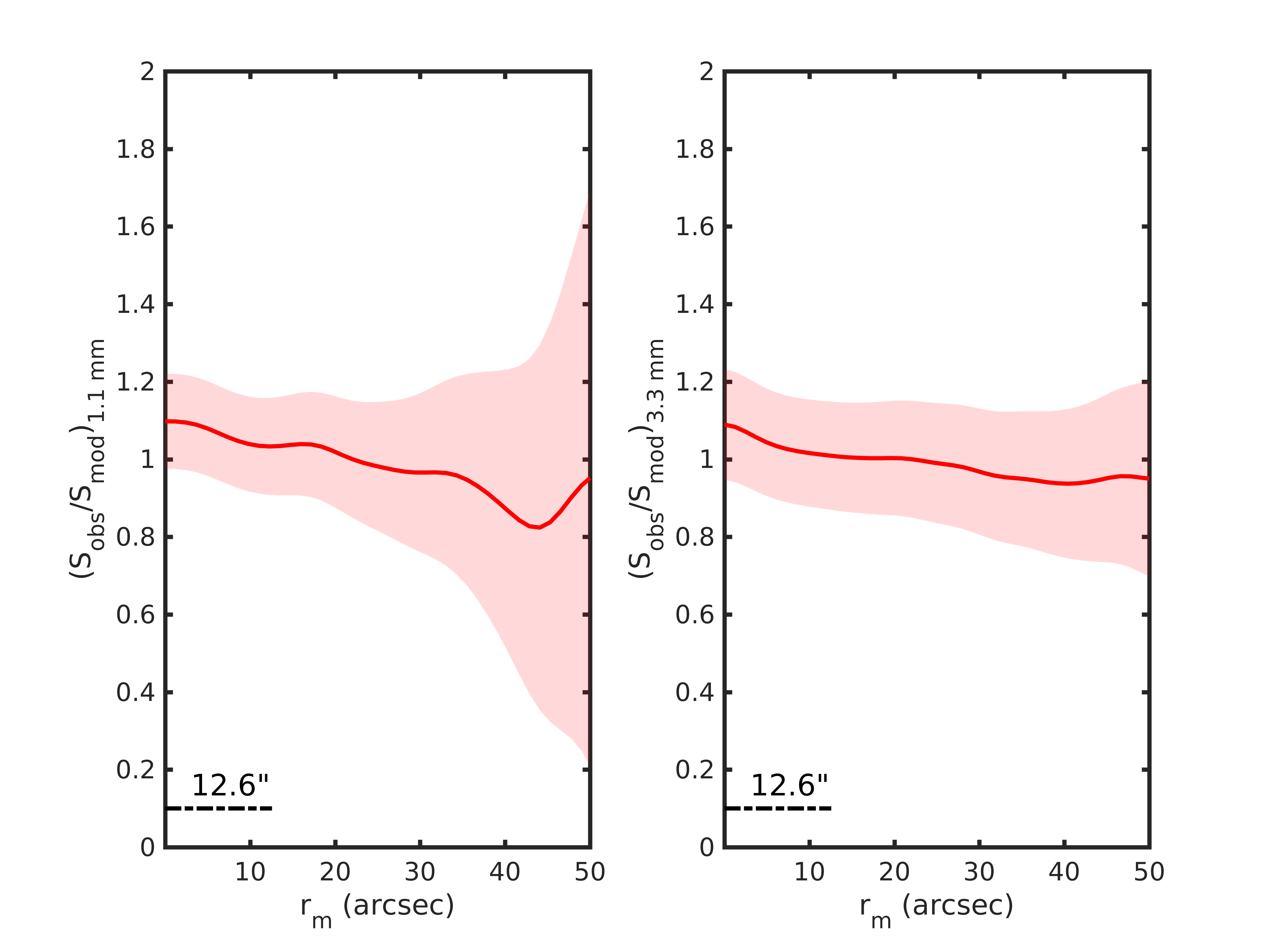}
\caption{Ratio between the observed emission profiles and the modeled emission profiles taking into account the radial opacity variations and new density and temperature profiles, as explained in Section \ref{new}. The shaded regions show the noise associated with the data.}\label{mm-final}
\end{figure}

Fig. \ref{mm-final} shows that the new density, temperature and opacity profiles can reproduce the observations. Although, as already discussed, these profiles are at the resolution of 12.6\arcsec, the model was considered at infinite (or intrinsic) resolution. We also check the results at the MUSTANG-2 resolution and find good agreement between the model and the observations (see Fig. \ref{mm-final-mustang}).

\section{Discussion}\label{discussion}
\subsection{New density and temperature profiles: comparison with previous profiles}

\begin{figure}
\includegraphics[width=0.5\textwidth]{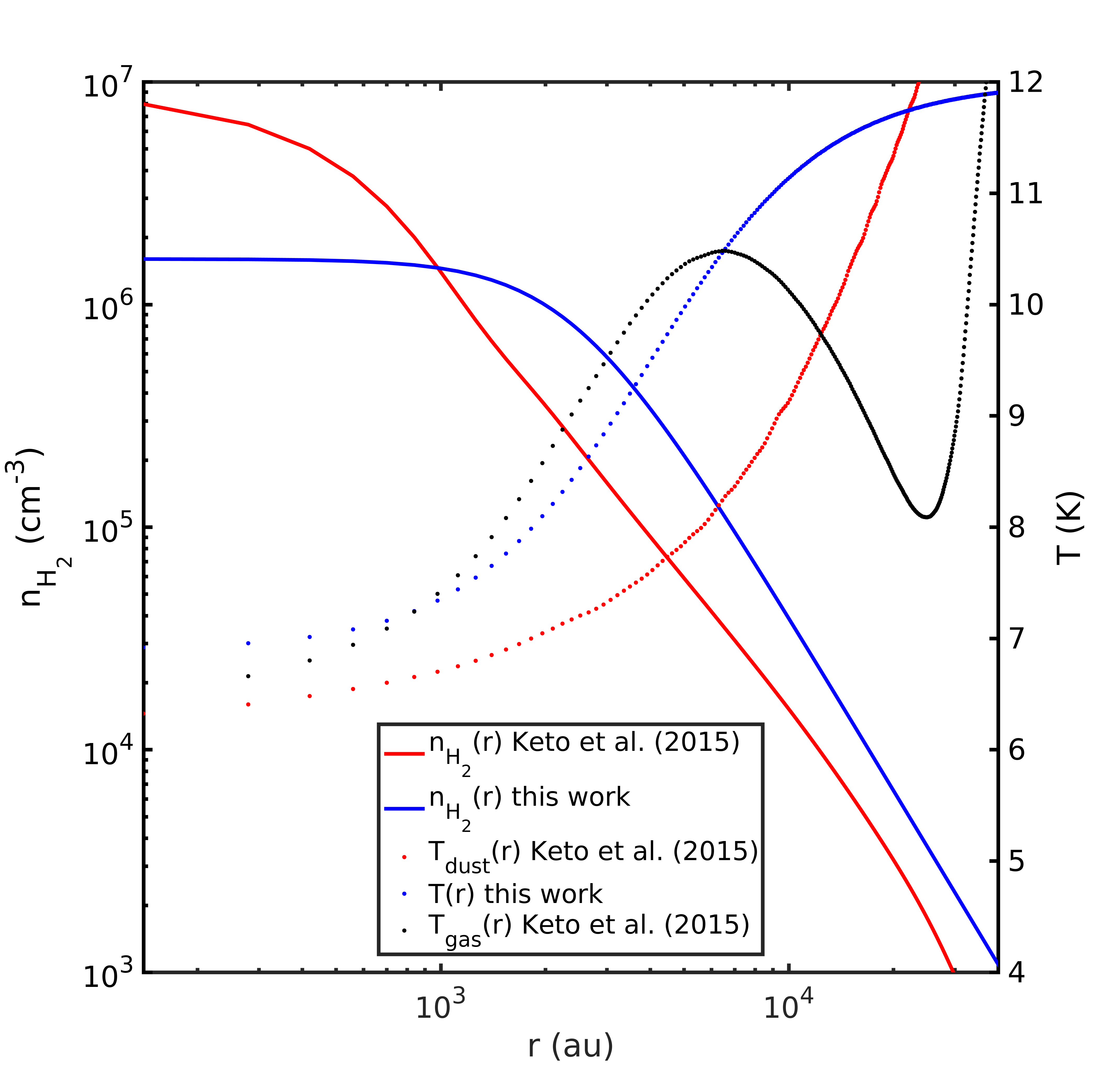}
\caption{New density (blue solid line) and temperature (blue dotted line) profiles derived as described in Section \ref{new}. In red, density (solid line) and dust temperature (dotted line) profiles from \citetads{2015MNRAS.446.3731K}. In black, gas temperature profile from \citetads{2015MNRAS.446.3731K}. }\label{model-keto}
\end{figure}

Fig. \ref{new_profiles} shows that the new density profile is flatter in the inner regions of the cloud, when compared to that deduced by \citetads{2007A&A...470..221C}. It also has a lower central density than the models from \citetads{2007A&A...470..221C} (by 25\%) and  \citetads{2015MNRAS.446.3731K} (by a factor of $\sim$5, see Fig. \ref{model-keto}). Nevertheless, this profile, as well as the one from  \citetads{2007A&A...470..221C}, gives a mass in the central $\sim$10\,000 au of 4 M$_{\sun}$, while the model from \citetads{2015MNRAS.446.3731K}, with a steep density gradient, gives a mass of 1.4 M$_{\sun}$.

The obtained dust temperature in the cloud center is approximately 1\,K higher than that derived by \citetads{2007A&A...470..221C} using also the equations from \citetads{2001A&A...376..650Z}, see their Fig. 4. The agreement with the model of \citetads{2015MNRAS.446.3731K} is fairly good. 

\subsection{Implication of opacity variations}\label{discussion_opacities}

Any variation of the opacity can arise from, for example, changes in the ice composition, grain coagulation or changes in the shape of the grains. In view of the fact that CO molecules are known to be frozen onto dust grains in the central parts of dense cores \citepads{1999ApJ...523L.165C}, the dust grains are expected both to increase in size and to change in composition, following the growth of icy mantles. 

The study from \citetads{1994A&A...291..943O} already predicted changes between the diffuse Interstellar Medium (ISM) and the densest parts of the core, including the effect of ice mantles growth. For example, at 1 mm, the change in opacity from the initial conditions, with no coagulation, to the dense regions with thick ice mantles is a factor of $\sim$3. Here, we find that from regions of density $\sim$10$^5$ cm$^{-3}$ to the central dense regions ($n_{\mathrm{H_2}}>10^6$ cm$^{-3}$), the opacity changes by a factor of $\sim$4. The opacity variations are consistent with the results from \citetads{1994A&A...291..943O} when ice mantle growth and coagulation are taken into account, but the values found here are systematically lower by a factor of 2. However, this difference is within the uncertainties \citepads{1994A&A...291..943O}, in addition to the effects of the beam, which dilutes the centrally concentrated structure and its emission, thus mimicking an opacity decrease. For example, if one compares the opacity at 3.3 mm at the resolution of 12.6\arcsec (Fig. \ref{kb-new}) with the opacity obtained at 9.7\arcsec resolution (Fig. \ref{kappa_3mm_intrinsic}), there are differences up to 20\%, thus bigger differences are expected between the actual structure and the one we observe. Nevertheless, the value of these opacities depend strongly on the initial density profile assumed, and they may vary as well if the external low density cloud and the filtering are considered. In fact, this also explains the fact that the opacity at the center of the core is very close to $\kappa_{1.2 \mathrm{mm}} = 0.005$ cm$^{2}$g$^{-1}$, which is the value used by \citetads{2007A&A...470..221C} for deriving the density profile through the 1.3 mm continuum observations of \citetads{1999MNRAS.305..143W}.

The variation at 3.3 mm, as already noted above, is heavily affected by the noise, and the increase of the opacity at r$\sim$7\,000\,au is also produced by the extended emission towards the north-east. However, as this is faint emission, if one considers a constant opacity at 3.3 mm with a value of $\sim$5$\times$10$^{-4}$ cm$^2$g$^{-1}$ the emission is well modeled within $\sim$20\% accuracy. Therefore, the variations at 3.3 mm may not be real or attributed to dust grains, but shows that more sensitive observations are needed.

\subsection{Comparison with a simple grain growth model}

As done in \citetads{2017A&A...606A.142C}, these results can be compared with a simple grain growth model applied to L1544. This model is based on the analytic estimation of grain growth presented by \citetads{2004PhRvL..93k5503B}, which was compared with the more complex grain growth model from \citetads{2009A&A...502..845O}, finding a good match. For a complete description of how this model is used,  see \citetads{2017A&A...606A.142C}. 

As discussed in \citetads{2017A&A...606A.142C}, models which do not account for a density evolution of a dense core, overestimate the grain sizes present in the core center by orders of magnitude. Since no information about the dynamical history is available for the model derived in Sect. \ref{new}, the cloud is assumed to follow the evolution described in \citetads{2015MNRAS.446.3731K}: we first obtain the ratio between the maximum density at each time and the central density value which best describes the current structure of L1544; this gives percentages which provide the time evolution of the central density. Such percentages are then applied to our new density profile, obtaining an approximation of the cloud evolution with time based on a Bonnor-Ebert sphere modeling. For the evolution of the size of the inner flatter region, Equation (2) from \citetads{2010MNRAS.402.1625K} was used. 

\begin{figure}
\includegraphics[width=0.5\textwidth]{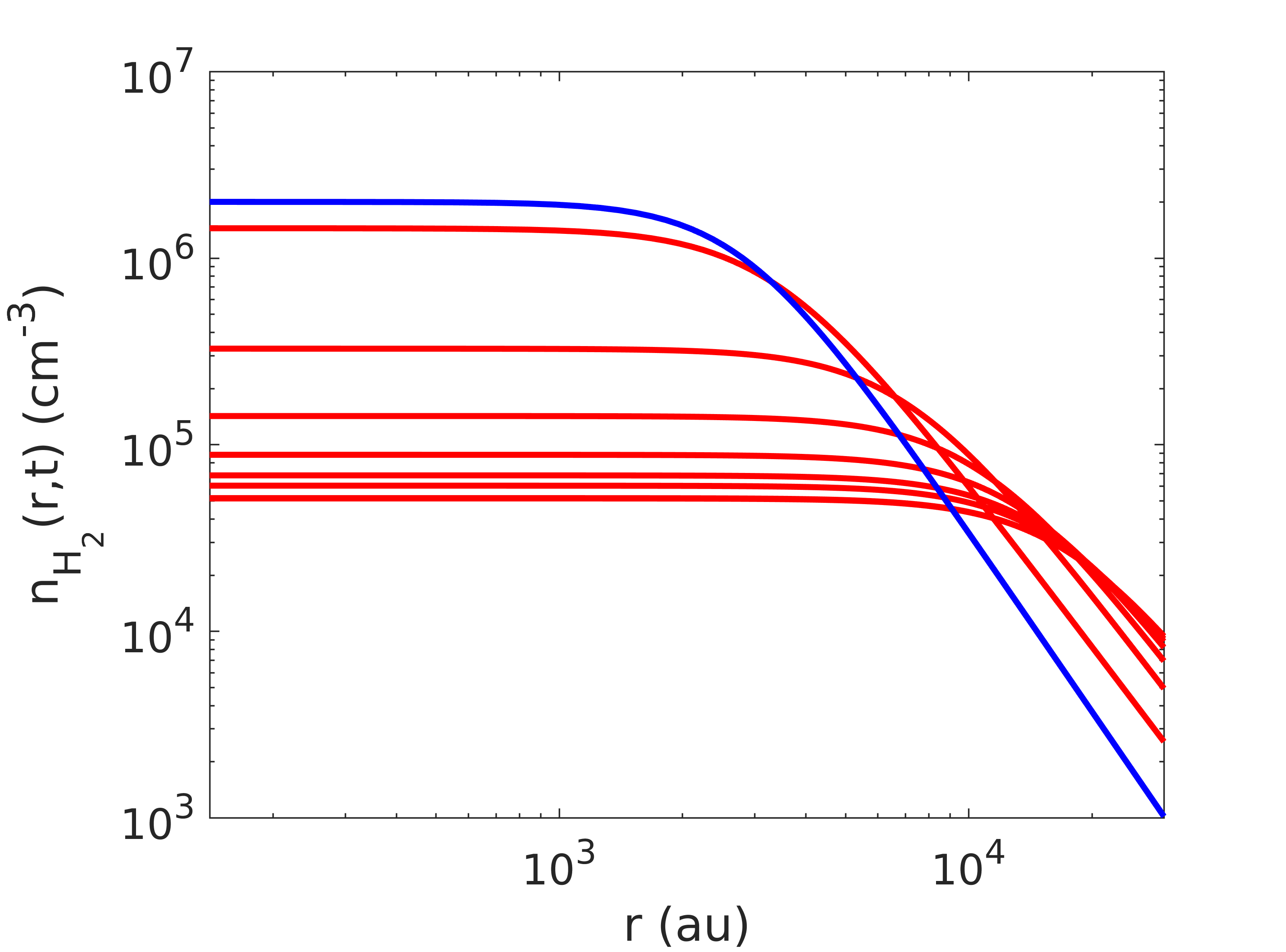}
\caption{Variation of the density with time extrapolated from that of \citetads{2015MNRAS.446.3731K}. The times at which the density is shown are, from low density to higher density: 0.06, 0.10, 0.60, 0.73, 0.85, 0.94, 1.01, 1.05 Myrs, and the current one, in blue, at t$\sim$1.06 Myrs.  }\label{density-time}
\end{figure}

This evolution of the density profile is seen in Fig. \ref{density-time}. We stress that this is an approximation only to estimate how much can grain growth impact in opacity variations at 1.1 and 3.3 mm.  

The final grain distribution found reaches sizes of $\sim$3-4 $\mu$m in the central 2\,000 au, which is consistent with the results from \citetads{2009A&A...502..845O} for densities of or above 10$^5$ cm$^{-3}$ and time evolution above 0.1 Myrs. However, transforming this to opacities using the code presented by \citetads{2016A&A...586A.103W}, in the same way as described in  \citetads{2017A&A...606A.142C}, yields very light changes in the opacity (less than 0.001\%), impossible to check observationally. However, this code does not include ice mantles, which could be the source of opacity change in the core: the closer to the center, the thicker the ice mantles. This needs to be checked with more complex and complete grain growth and opacity models, as well as higher angular resolution observations (e.g. ALMA observations).

\section{Conclusions}\label{conclusions}

We have used two new millimeter facilities, AzTEC at the LMT and MUSTANG-2 at the GBO, to study the physical structure and dust emission properties of the pre-stellar core L1544. Their sensitivity and resolution have allowed us to study the inner $\sim$1700 au of the core with a beam of 12.6\arcsec.
The results show that previous density profiles deduced by \citetads{2015MNRAS.446.3731K} and \citetads{2007A&A...470..221C} are not able to explain the emission of the core at 1.1 and 3.3 mm without invoking dust opacity variations, which in turn implies a need to re-determine the physical structure of the core, as these models did not consider opacity variations.

Although future work including the emission at different wavelengths is needed, we modified self-consistently the model from \citetads{2007A&A...470..221C} making use of the Abel transform and the analytical formulae from \citetads{2001A&A...376..650Z}. We thus obtained a new density and temperature profile, together with the radial opacity variations seen towards the core. These opacity gradients show increasing opacities toward the core center, where thick icy mantles are expected. However, the measured opacities are about a factor of 2 lower than those for coagulated dust grains \citepads{1994A&A...291..943O}. Our model predicts dust grains of 3-4 $\mu$m in size within the central 2\,000 au of L1544, which indicates that dust coagulation may not be affecting the emission at millimeter wavelengths yet. 

This study needs to be expanded to a sample of cores, in order to test the general validity of these conclusions. Furthermore, interferometric observations are needed to study the yet unresolved center of the core. These will allow quantitative comparison between observations and grain growth models, which predict relatively large (3-4 $\mu$m) dust grains, in the central 2\,000 au of L1544. 

\begin{acknowledgements}
 The authors thank the anonymous referee for the careful
reading and useful comments. The Green Bank Observatory is a facility of the National Science Foundation operated under cooperative agreement by Associated Universities, Inc. RAG's participation in this project was supported by NSF grant AST 1636621 in support of TolTEC, the next generation mm-wave camera for LMT. The AzTEC instrument was built and operated through support from NSF grant 0504852 to the Five College Radio Astronomy Observatory. The authors gratefully acknowledge the many contributions of David Hughes in leading the LMT to its successful operational state. The authors also thank the personnel and observers at the LMT that helped obtain the observations, including Alyssa Sokol, Carolina Rodr\'iguez, Marcos Emir Moreno, and Emmaly Aguilar. ACT, PC, and JEP acknowledge the financial support of the European Research Council (ERC; project PALs 320620).

\end{acknowledgements}
\bibliography{biblio}
\bibliographystyle{aa}

\newpage
\begin{appendix}
\section{Transfer functions}\label{transfer_function}
\begin{figure}[h]
\includegraphics[width=0.5\textwidth]{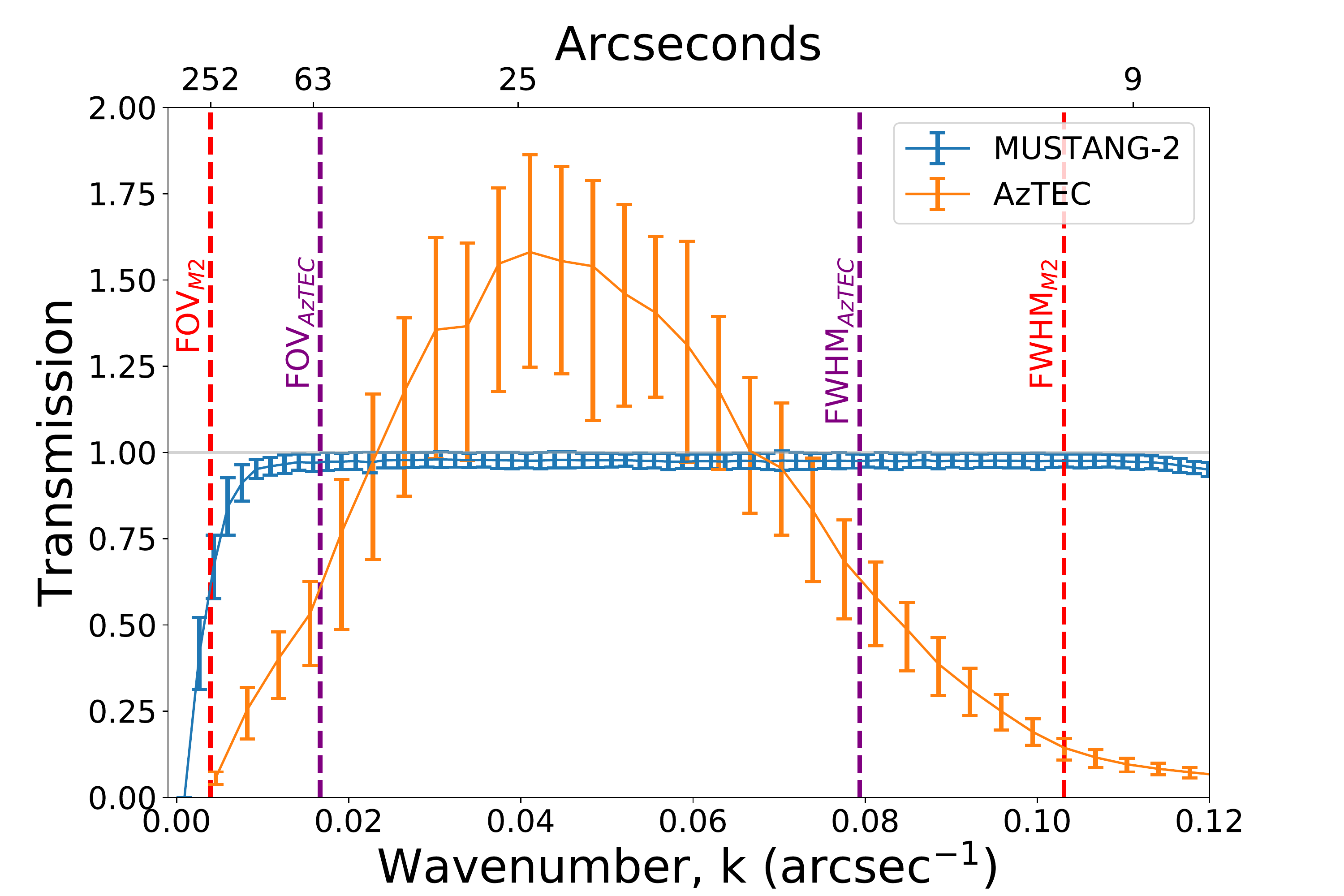}
\caption{Transfer functions for MUSTANG-2 (in blue) and AzTEC (in orange). Recovery of the smallest angular scales is limited by the full width at half maximum (FWHM) beam size of each instrument, indicated on the right with purple dashed lines for AzTEC and red dashed lines for MUSTANG-2. Both instruments subtract a common mode component to remove atmospheric noise, resulting in the loss of angular scales larger than the field of view (FoV), which is indicated by the dashed lines on the left. In addition, the AzTEC data reduction process includes an adaptive Wiener filter to ensure the recovery of point source fluxes with the correct amplitudes, which results in the amplitude of the transfer functions exceeding unity at some angular scales. For MUSTANG-2, the FoV and the FWHM are further apart and this filtering step was not required. }\label{figure_transfer_function}
\end{figure}

We derived the transfer functions of the pipelines used for AzTEC and MUSTANG-2, in order to verify the spatial scales we are recovering with these observations. Both transfer functions are shown in Fig. \ref{figure_transfer_function}.

\section{Comparison with NIKA}\label{comparison_nika}

\begin{figure}[h]
\includegraphics[width=0.45\textwidth]{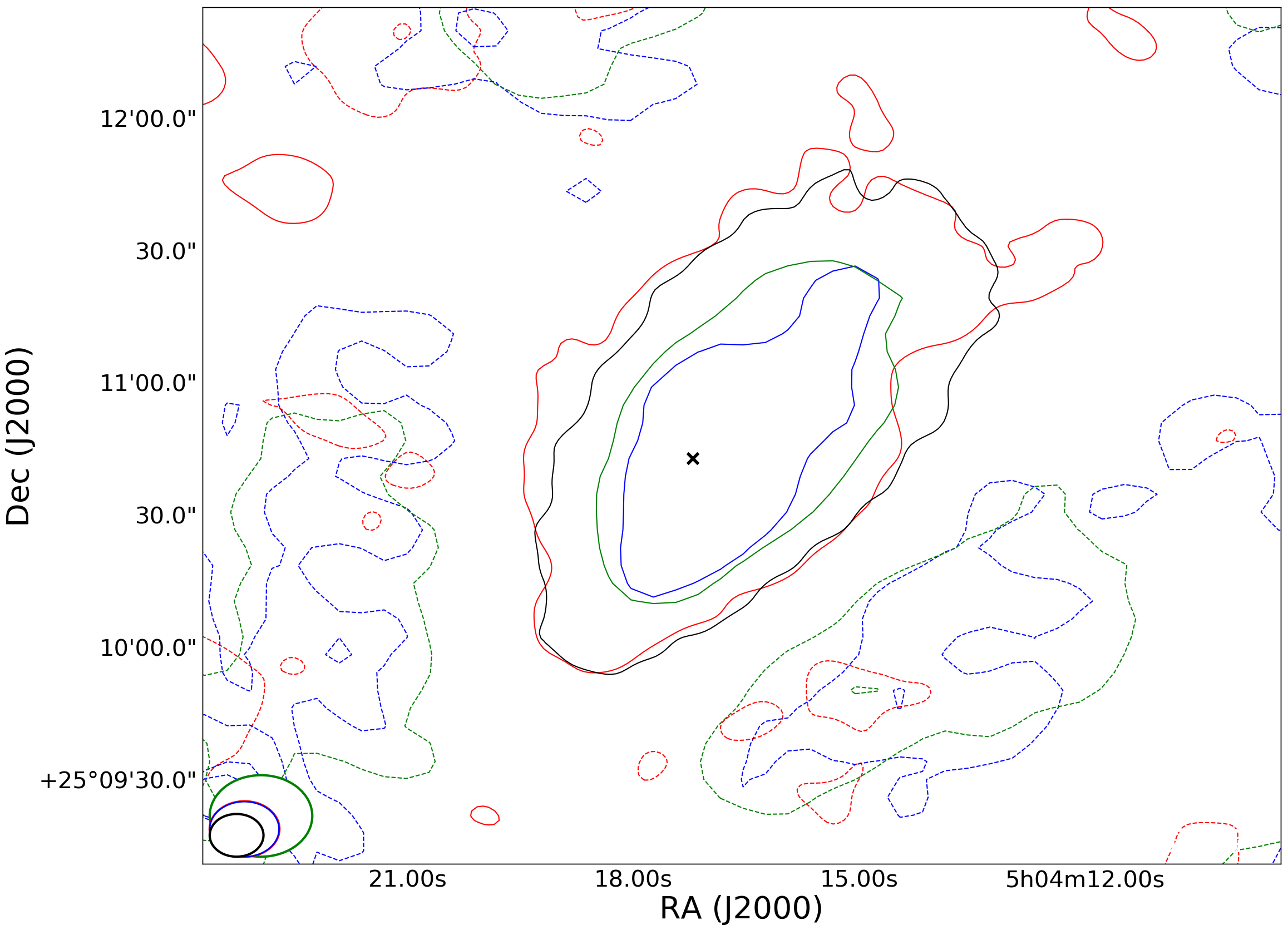}
\caption{Contour map of the emission of L1544 observed with NIKA at 1.25 mm (in blue) and 2 mm (in green) by \citetads{2017A&A...606A.142C}, together with that observed with AzTEC at 1.1 mm (in red) and MUSTANG-2 at 3.3 mm (in black). The corresponding solid lines follow each 3$\sigma$ contour, while the dashed lines follow the negative bowls formed due to filtering, indicating a -10\% of the peak emission of each map. The HPBWs are on the bottom left corner of the figure. }\label{nika_contour}
\end{figure}

To check the differences of the filtering effects on the continuum millimeter maps, Fig. \ref{nika_contour} shows the contours of the emission seen with NIKA, AzTEC and MUSTANG-2. NIKA suffers from more filtering than AzTEC, as it shows more extended negative bowls than the AzTEC map. On the other hand, MUSTANG-2 is the instrument that shows the least filtering as expected, since it is the camera with the largest field of view.

\section{Spectral Energy Distribution fit}\label{sed_fit_appendix}
\begin{figure}[h]
\includegraphics[width=0.45\textwidth]{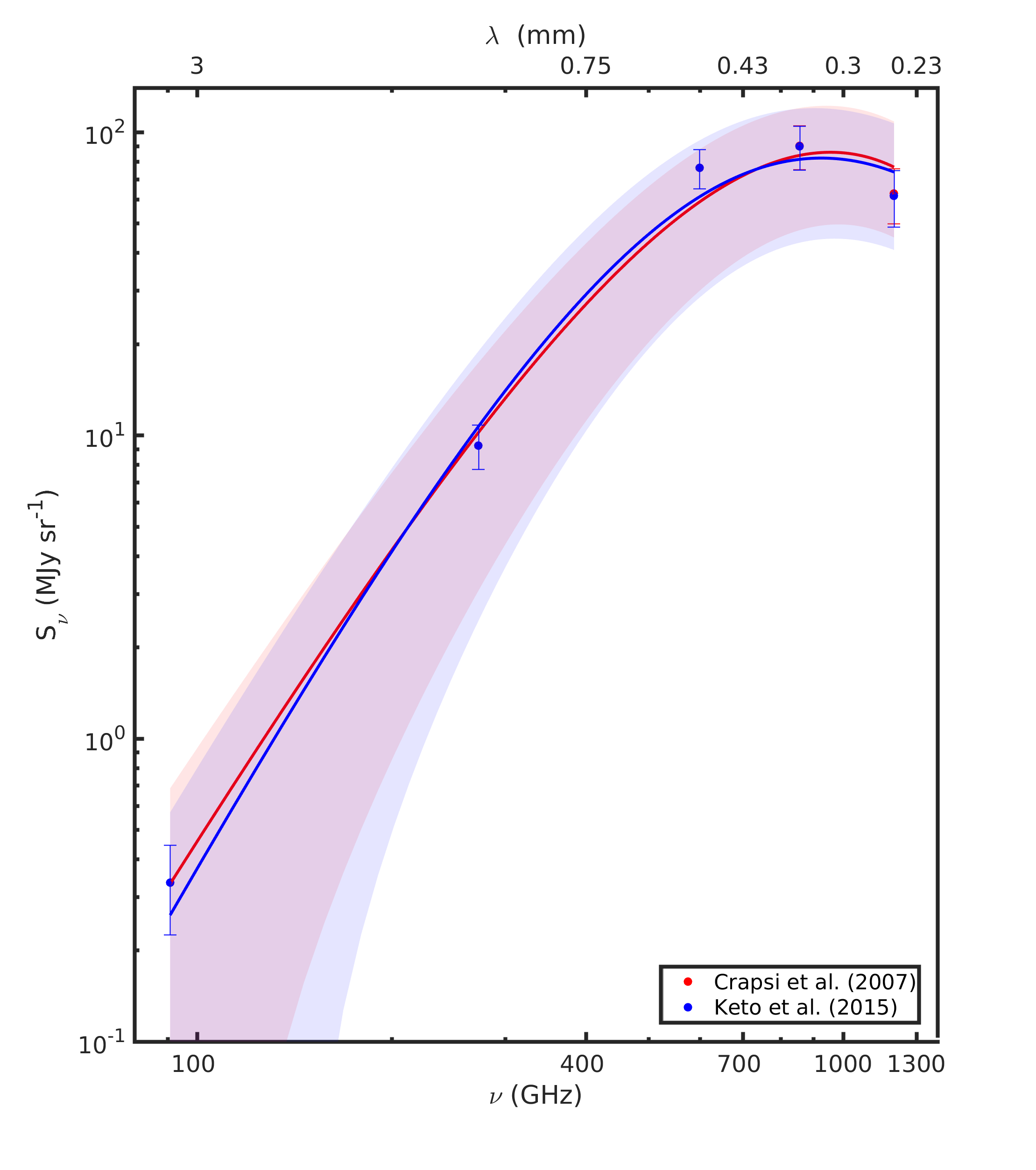}
\caption{Fit of the SED to the 5 spectral windows available from \textit{Herschel}/SPIRE, AzTEC and MUSTANG-2 at 250, 350, 500, 1100, and 3300 $\mu$m towards the central pixel after smoothing and regridding all maps to the resolution of the 500$\mu$m band, using the \citetads{2015MNRAS.446.3731K} model in blue, and the \citetads{2007A&A...470..221C} model in red. The shaded regions indicate the 95\% confidence intervals of the fitted parameters. The errorbars indicate the weights used in the fitting.}\label{SED_fit}
\end{figure}

\section{Modeling based on previous results}\label{modeling_appendix}
\begin{figure}[h]
\includegraphics[width=0.5\textwidth]{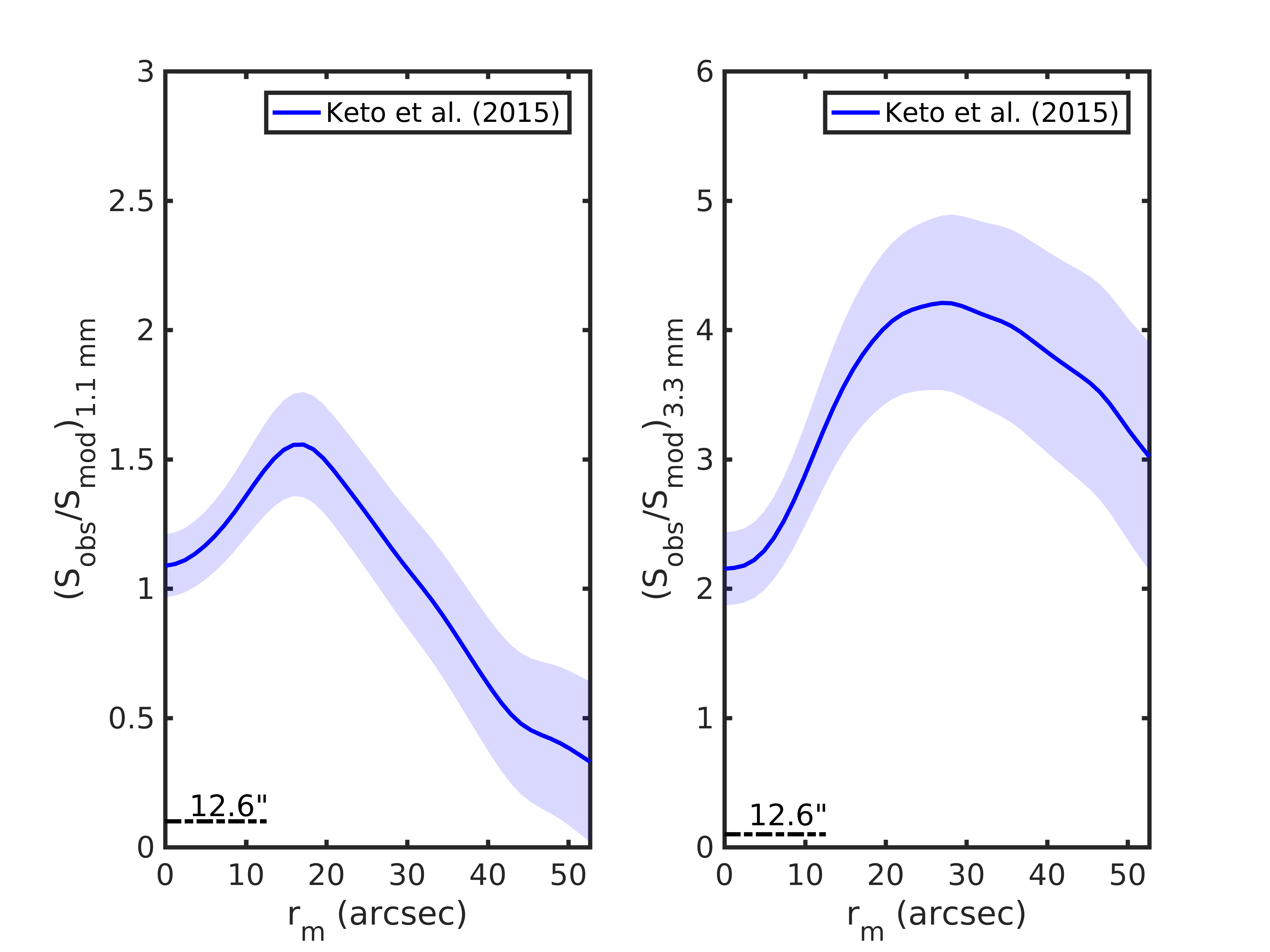}
\caption{Ratio between the observed emission profiles and the modeled emission profiles using the same modeling parameters as in \citetads{2017A&A...606A.142C}. The shaded regions show the noise associated with the data. }\label{mm-old}
\end{figure}

Here we compare the observations with the modeled emission using the physical structure from \citetads{2015MNRAS.446.3731K}, following the same steps as in Section \ref{modeling1}, but using the spectral index and the opacity values from \citetads{2017A&A...606A.142C}, which are $\kappa_{250\mathrm{\mu m}} = 0.2$ cm$^2$g$^{-1}$ and $\beta=2.3$. This is to check the results following exactly the same procedure of \citetads{2017A&A...606A.142C}. The results are presented in Fig. \ref{mm-old}. As it is seen, the model does not reproduce the observations.

\section{$A_V$ from Herschel/SPIRE}\label{av_herschel}

To derive a value for the visual extinction, $A_{V}$, corresponding to the external low density layer of the cloud, we take the $N_{\mathrm{H_2}}$ map from \citetads{2016A&A...592L..11S}, derived from the emission of Herschel/SPIRE, and transformed it into $A_{V}$ using the following equation \citepads{1978ApJ...224..132B}:

\begin{equation}
N_{\mathrm{H_2}} = 9.4\times10^{20} \mathrm{cm}^{-2} (A_V/\mathrm{mag}).
\end{equation}

\begin{figure}[h]
\includegraphics[width=0.5\textwidth]{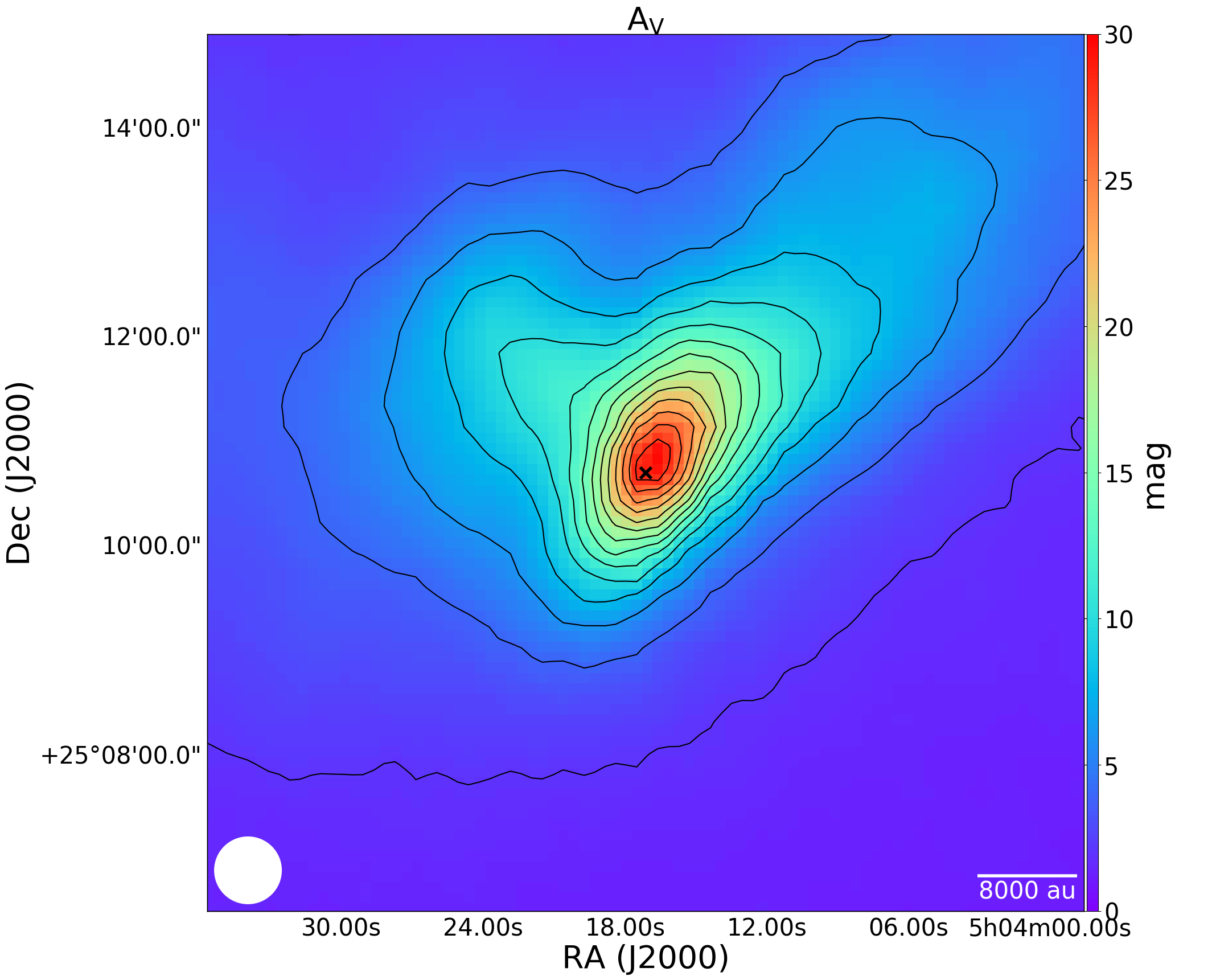}
\caption{Visual extinction map of L1544 derived from the $N_{\mathrm{H_2}}$ map from \citetads{2016A&A...592L..11S}. The contours represent steps of 2 magnitudes.}\label{av}
\end{figure}

This produces the $A_V$ map presented in Fig. \ref{av}. At distances far away from the dust continuum peak ($>$10\,000 au), the $A_V$ has a value of $\sim$2 mag. This is not seen in the ground based millimeter maps as this part of the cloud is filtered out. This outer layer shields the inner core structure seen in our data from the external radiation field, and therefore it affects the internal temperature.

\section{A different type of profile}\label{density_option}

Here we consider another starting point for the derivation of the density, temperature and opacity profiles that reproduce our observations.

From the Abel transform, one direct measurement is the factor $\kappa_{\nu}n_{\mathrm{H_2}}$. Therefore we do the following:

\begin{enumerate}
\item $\kappa_{1.1\mathrm{mm}}n_{\mathrm{H_2}}$ and  $\kappa_{3.3\mathrm{mm}}n_{\mathrm{H_2}}$ are derived from the Abel transform (Eq. \eqref{abel}) for each band.
\item $\beta$ is derived from $\beta = \frac{\log(A_1/A_3)}{\log(\nu_{\mathrm{1.1mm}}/ \nu_{\mathrm{3.3mm}})}$ , being $A_1$ and $A_3$ the right hand side of Eq. \eqref{abel} at 1 and 3 mm, respectively.  
\item The density is derived from $\kappa_{3.3\mathrm{mm}}n_{\mathrm{H_2}}$, assuming $\kappa_{3.3\mathrm{mm}}$ constant across the cloud and consistent with \citetads{1994A&A...291..943O} thin ice mantles and $\beta$ = 2.0, i.e. $\kappa_{3.3\mathrm{mm}}$=0.0011 cm$^2$g$^{-1}$. 
\item The variation of the opacity at 1.1 mm is derived from $\kappa_{1.1\mathrm{mm}}n_{\mathrm{H_2}}$.
\item The temperature profile is derived for the opacity and spectral index variation at 1.1 mm. 
\end{enumerate}

\begin{figure}[h]
\includegraphics[width=0.5\textwidth]{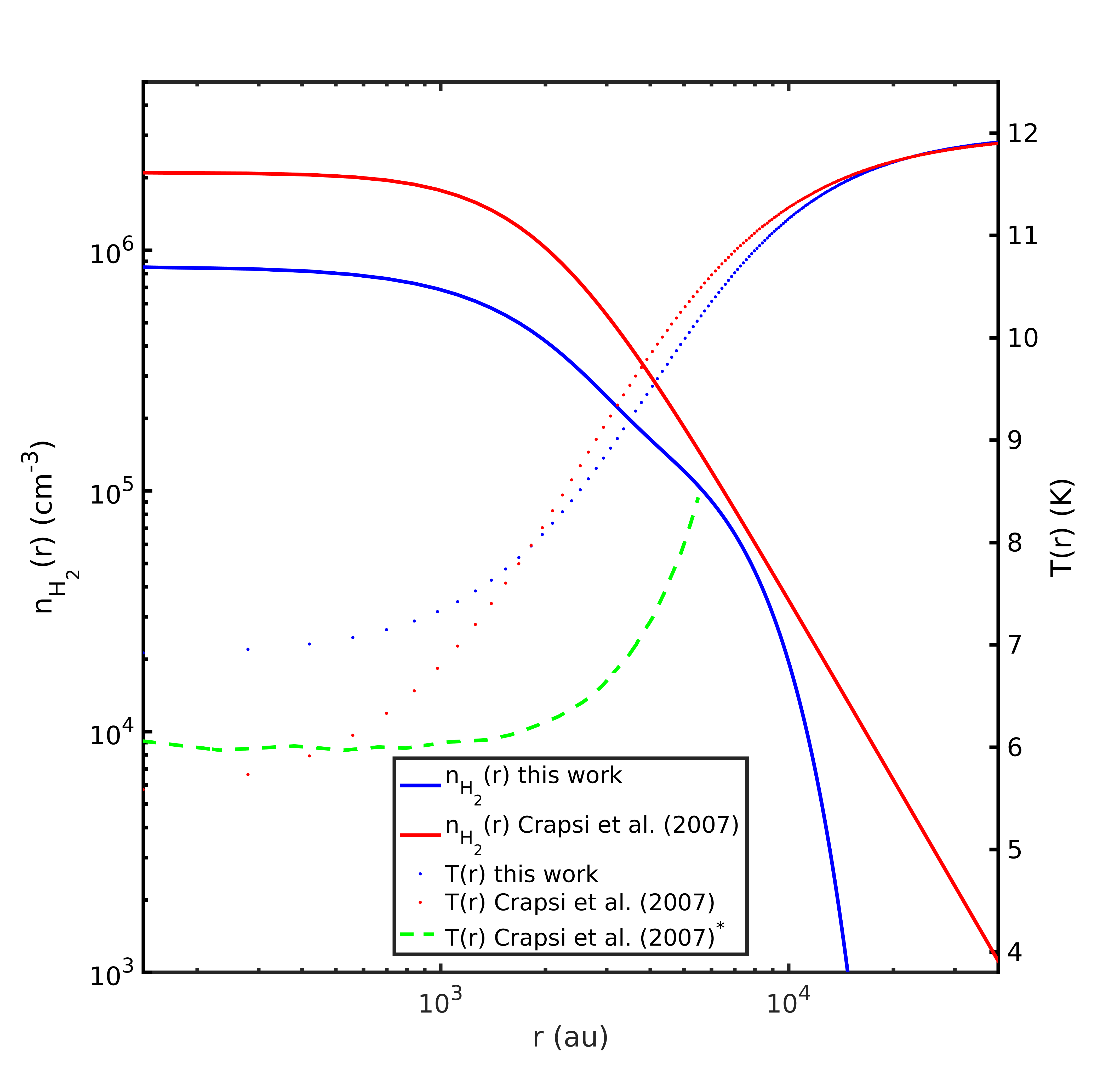}
\caption{New density (blue solid line) and temperature (blue dotted line) profiles derived as described in Section \ref{density_option}. In red, density (solid line) and temperature (dotted line) profiles from \citetads{2007A&A...470..221C}. In green, temperature profile from \citetads{2007A&A...470..221C} derived using the formulae from \citetads{2001A&A...376..650Z}, here \citetads{2007A&A...470..221C}$^{\ast}$. }\label{alternative_profiles}
\end{figure}

\begin{figure}[h]
\includegraphics[width=0.5\textwidth]{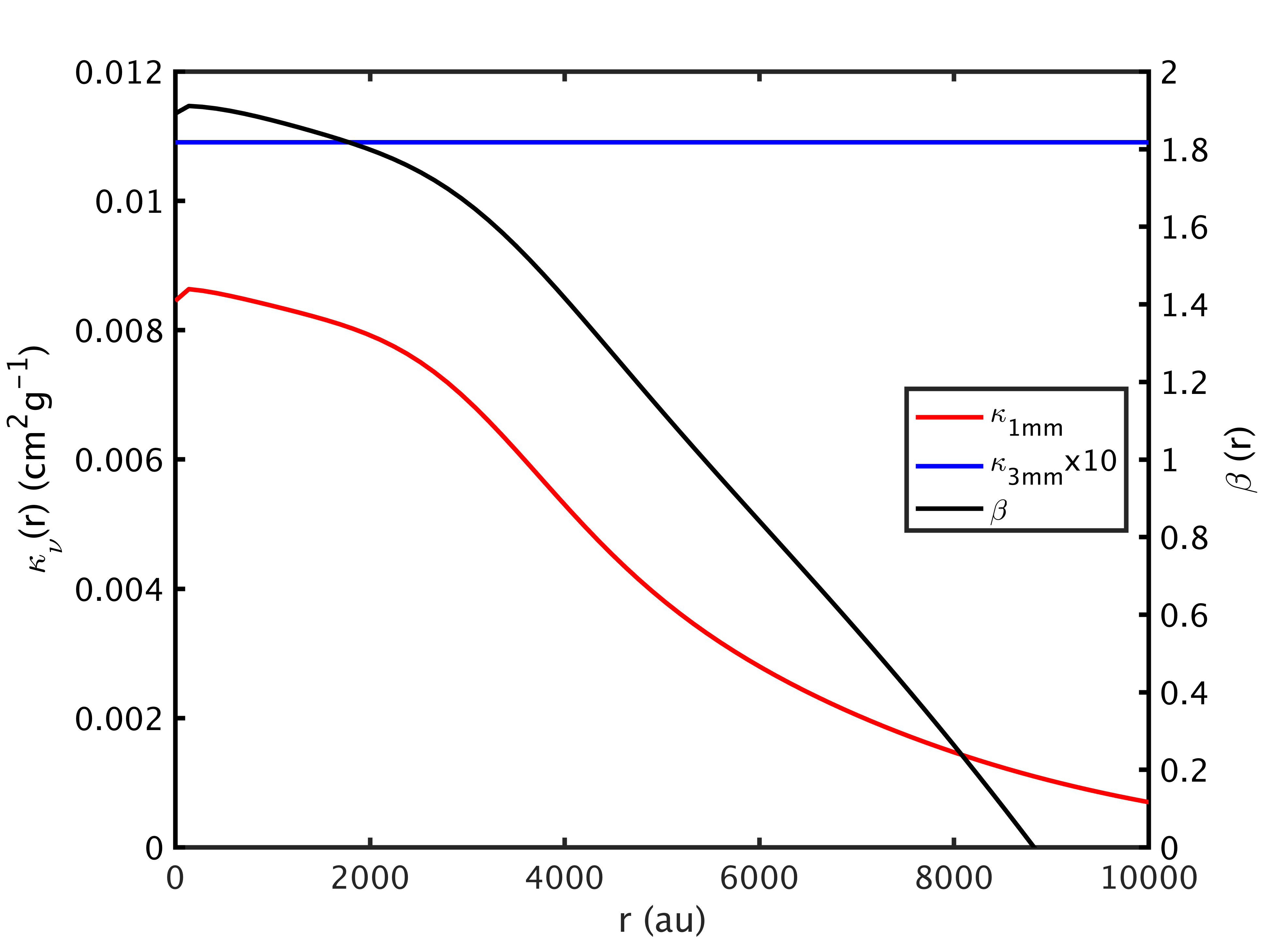}
\caption{Radial opacity and spectral index variations derived as described in Section \ref{density_option}. }\label{alternative_opacity}
\end{figure}

\begin{figure}[h]
\includegraphics[width=0.5\textwidth]{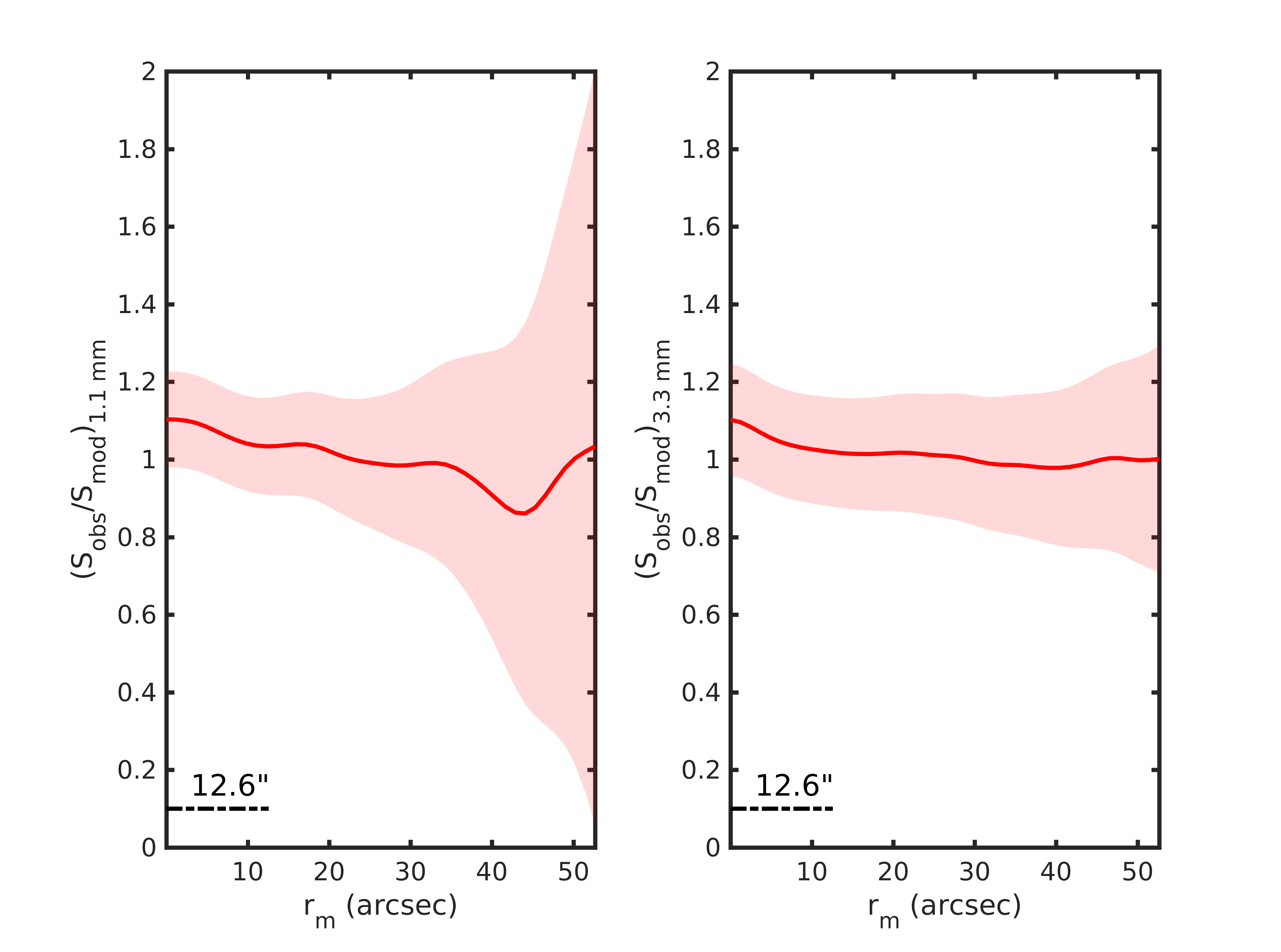}
\caption{Ratio between the observed emission profiles and the modeled emission profiles taking into account the radial opacity variations, and new density and temperature profiles, as explained in Section \ref{density_option}. The shaded regions show the noise associated with the data.}\label{mm-final_alternative}
\end{figure}

Iterating the previous points until the opacity converges, we find the new density, temperature and opacity profiles in Fig. \ref{alternative_profiles} and Fig. \ref{alternative_opacity}. These results  reproduce the observations, as shown in Fig. \ref{mm-final_alternative}. 

The opacities in this case are similar to those presented by \citetads{1994A&A...291..943O} for thin ice mantles. However, the central density has declined below 10$^6$ cm$^{-3}$. This central density is inconsistent with previous observations (e.g. \citeads{2005ApJ...619..379C}) and modeling including simple chemistry and dynamics constrained by observed line profiles \citepads{2010MNRAS.402.1625K}. We therefore follow the method presented in the main text, but ask the reader to be cautious, as more observations are needed to constrain the absolute values.   

\section{MUSTANG-2 resolution}

\begin{figure}[h]
\includegraphics[width=0.5\textwidth]{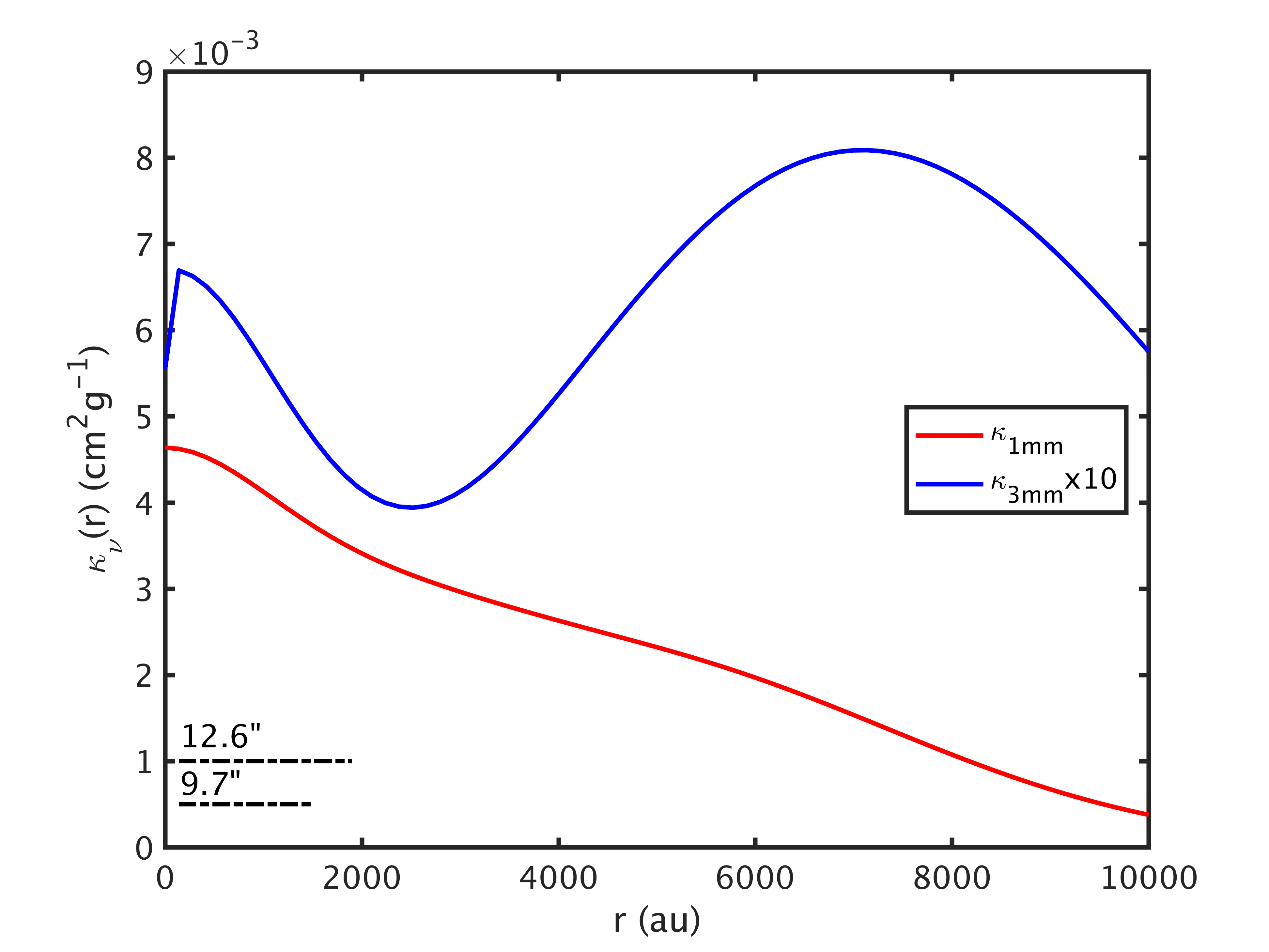}
\caption{Radial opacity variations derived as described in Section \ref{new}. The resolution of the opacity at 1.1 mm is 12.6\arcsec, and the one at 3.3 mm is 9.7\arcsec, both shown in the figure.}\label{kappa_3mm_intrinsic}
\end{figure}

\begin{figure}[h]
\includegraphics[width=0.5\textwidth]{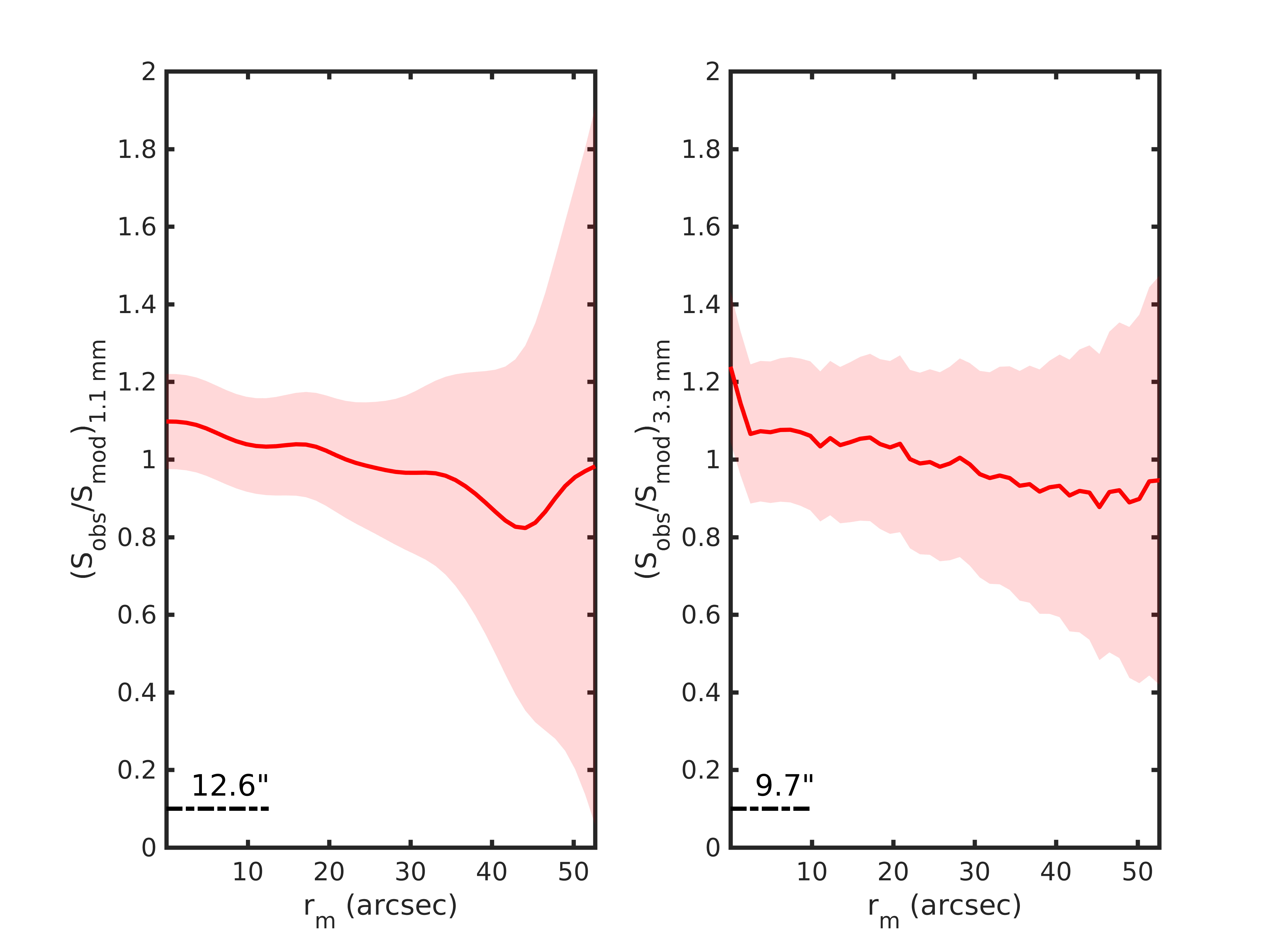}
\caption{Ratio between the observed emission profiles and the modeled emission profiles taking into account the radial opacity variations and new density and temperature profiles, as explained in Section \ref{new}. The shaded regions show the noise associated with the data, and the bars the resolution of the data.}\label{mm-final-mustang}
\end{figure}

Fig. \ref{kappa_3mm_intrinsic} shows the radial opacity variations derived at the intrinsic resolution of each map (9.7\arcsec at 3.3 mm and 12.6\arcsec at 1.1 mm), and Fig. \ref{mm-final-mustang} shows the corresponding comparison between the observed and the modeled emission. The opacity at 3.3 mm at 9.7\arcsec resolution shows differences up to 20\% when compared to the same opacity derived at a resolution of 12.6\arcsec.

\end{appendix}

\end{document}